%% file: main.tex
\documentclass[a4paper,11pt]{article}
\pdfoutput=1 

\usepackage{jcappub} 
\usepackage[T1]{fontenc}
\usepackage{graphicx}
\usepackage[utf8]{inputenc}
\usepackage[english]{babel} 
\usepackage{enumitem}
\usepackage{float}
\usepackage{amssymb}
\usepackage{amsmath}
\usepackage{slashed}
\usepackage{color}
\usepackage{mathtools}
\usepackage{multirow}
\usepackage{hyperref}
\usepackage{subfig}
\usepackage{makecell}
\usepackage{cancel}
\def\baredth{\;\overline{\raise1.0pt\hbox{$'$}\hskip-6pt
		\partial}\;}
\def\edth{\;\raise1.0pt\hbox{$'$}\hskip-6pt\partial\;}

\newcommand{\kafzero}{(k_{AF})_{{0}}}

\newcommand{\bkafzero}{(\bar{k}_{AF})_{{0}}}

\newcommand{\calK}{{\mathcal K}}

\newcommand{\calZ}{{\mathcal Z}}
\newcommand{\wt}[1]{\widetilde{#1}}

\newcommand{\cambcpt}{{\tt camb-cpt}}

\input{Planck}
\input{shortcuts}

\title{\boldmath Probing Lorentz-violating electrodynamics with CMB polarization}

\author[a,b]{L. Caloni,}
\author[a,b,c]{S. Giardiello,}
\author[a,b]{M. Lembo,}
\author[b,a]{M. Gerbino,}
\author[d,e]{G. Gubitosi,}
\author[b,a]{M. Lattanzi,}
\author[a,b,g]{and L. Pagano}

\affiliation[a]{Dipartimento di Fisica e Scienze della Terra, Università degli Studi di Ferrara, via Saragat 1, I-44122 Ferrara, Italy}
\affiliation[b]{Istituto Nazionale di Fisica Nucleare, Sezione di Ferrara, via Saragat 1, I-44122 Ferrara, Italy}
\affiliation[c]{School of Physics and Astronomy, Cardiff University, The Parade, CF24 3AA Cardiff, Wales, UK}
\affiliation[d]{Dipartimento di Fisica Ettore Pancini, Universit\`a di Napoli “Federico II”, Complesso Univ. Monte S. Angelo, I-80126
Napoli, Italy}
\affiliation[e]{Istituto Nazionale di Fisica Nucleare, Sezione di Napoli, Complesso Univ. Monte S. Angelo, I-80126 Napoli, Italy}
\affiliation[g]{Institut d'Astrophysique Spatiale, CNRS, Univ. Paris-Sud, Universit\'{e} Paris-Saclay, B\^{a}t. 121, 91405 Orsay cedex, France}

\emailAdd{luca.caloni@unife.it}
\emailAdd{serena.giardiello@unife.it}
\emailAdd{margherita.lembo@unife.it}
\emailAdd{giulia.gubitosi@unina.it}
\emailAdd{gerbino@fe.infn.it}
\emailAdd{lattanzi@fe.infn.it}
\emailAdd{luca.pagano@unife.it}

\abstract{We perform a comprehensive study of the signatures  of Lorentz violation in electrodynamics on the Cosmic Microwave Background (CMB) anisotropies. In the framework of the minimal Standard Model Extension (SME), we consider effects generated by renormalizable operators, both CPT-odd and CPT-even. These operators are responsible for sourcing, respectively, cosmic birefringence and circular polarization. 
We propagate jointly the effects of all the relevant Lorentz-violating parameters to CMB observables and provide constraints with the most recent CMB datasets.
We bound the CPT-even coefficient to $k_{F,E+B} < 6.02 \times 10^{-32}$ at 95\% CL. This improves previous CMB bounds by more than one order of magnitude.
The limits we obtain on the CPT-odd coefficients, i.e. $|k_{(V)00}^{(3)}| < 1.05 \times 10^{-44} \; {\rm GeV}$ and $|\mathbf{k_{AF}}| < 0.51 \times 10^{-44} \; {\rm GeV}$ at 95\% CL, are respectively one and two orders of magnitude stronger than previous CMB-based limits, superseding also bounds from non-CMB searches. 
This analysis provides the strongest constraints to date on CPT-violating coefficients in the minimal SME from CMB searches.
}

\begin{document}
\maketitle
\flushbottom

\section{Introduction}

Lorentz symmetries are at the foundation of the current description of nature. However, theoretical investigations have suggested that they may only be exact symmetries at low energies \cite{Addazi:2021xuf, Amelino-Camelia:2008aez, Mattingly:2005re}. Motivations for this hypothesis are rooted in quantum gravity. Therefore, it is expected that the energy scale at which Lorentz invariance could be violated is the Planck scale. While the magnitude of this scale might discourage searches for Lorentz violations,  
high-precision experimental tests might be sensitive to their small low-energy residual effects.

There are currently a number of different theoretical frameworks describing departures from Lorentz symmetries \cite{Addazi:2021xuf, Amelino-Camelia:2008aez, Mattingly:2005re}. The most conservative approach is that of effective field theory, which incorporates  Lorentz violation via the introduction of extra tensors in the Lagrangian of the standard model. The new operators can be ordered according to their mass dimension: operators which introduce Lorentz violations at some high-energy scale have mass dimension higher than four, and therefore are non-renormalizable. Without some custodial symmetries, these operators might also induce Lorentz violations in operators with lower mass dimension. The lower-dimension renormalizable operators produce effects that are not suppressed by the high-energy scale, and could in principle dominate over the non-renormalizable operators, possibly leading to stronger signatures on low-energy physics.
The Lagrangian containing such renormalizable terms, known as minimal Standard Model Extension (SME), was first derived in \cite{Colladay:1998fq}. 
In this work we  focus on the radiation sector of the SME Lagrangian \cite{Kostelecky:2002hh} and test Lorentz invariance with observations of the CMB. We concentrate on renormalizable operators, leaving the study of the non-renormalizable operators \cite{Kostelecky:2009zp} to a future work.

The Cosmic Microwave Background (CMB) is an ideal probe of possible departures from standard electrodynamics. The CMB radiation is linearly polarized due to Compton scattering at the epochs of recombination and reionization \cite{Hu:2001bc}. 
A non standard propagation of light might induce distinctive patterns on the CMB polarization.
A very well known example is the cosmic birefringence effect, namely the in-vacuo rotation of the linear polarization plane of the CMB radiation\footnote{Faraday rotation induced by the interaction of the CMB with primordial magnetic fields can also produce a rotation of the CMB polarization which is proportional to the square of the radiation wavelength, see e.g., Refs.~\cite{Kosowsky:1996yc, Scoccola:2004ke, Campanelli:2004pm}. We do not consider Faraday rotation in this work.}.

Since the CMB last scattering surface is the farthest source of electromagnetic radiation available in nature, the cosmic birefringence effect accumulates during propagation of the CMB, increasing the chances of detecting a non-vanishing signal.
Evidence for new physics could also come from the observation of a sizable level of circular polarization. In the standard cosmological model, circular polarization is not expected at the time of last scattering, even though a tiny amount can be generated by known physics at a later time \cite{Kosowsky:1996yc,Cooray:2002nm,Giovannini:2009ru,De:2014qza,Montero-Camacho:2018vgs,Lemarchand:2018lfy,Ejlli:2018ucq} as CMB photons propagate across the Universe.  

As we will show in this paper, some combinations of these effects are expected within the SME framework, depending on which operators are considered. 
Among the two operators analyzed in our work, one violates CPT symmetry and is responsible for the generation of cosmic birefringence. The CPT-even operator, instead, leads to the generation of circular polarization from the conversion of the primordial linear polarization components.
While we will focus on the SME framework, we remark that both birefringence and the generation of circular polarization can emerge in other theoretical scenarios. In particular, cosmic birefringence  can be generated by Chern-Simons terms in the electrodynamics Lagrangian \cite{Carroll:1989vb, Carroll:1998zi, Li:2008tma, Pospelov:2008gg}, by the coupling of the electromagnetic field to quintessence \cite{Giovannini:2004pf, Balaji:2003sw, Liu:2006uh} or axion \cite{Finelli:2008jv} fields, or in quantum-gravity motivated effective theories for electromagnetism \cite{Myers:2003fd, Gubitosi:2009eu, Gubitosi:2010dj}. These scenarios might be distinguished because they predict different dependence on the frequency of the CMB signal \cite{Gubitosi:2012rg, Galaverni:2014gca} and on the propagation direction \cite{Kamionkowski:2008fp, Pospelov:2008gg, Li:2008tma, Gubitosi:2010dj, Caldwell:2011pu}.
Production of circular polarization is instead predicted by several scenarios beyond the standard model of particle physics, including a possible coupling between photons and an external vector field via a Chern-Simons term~\citep{Alexander:2008fp}, the Cotton-Mouton effect~\citep{Ejlli:2016avx}, propagation of CMB photons in a non-commutative spacetime~\citep{Tizchang:2016vef} and other non-standard effects~\citep{Kostelecky:2007zz,Ejlli:2017uli,Inomata:2018vbu,Bartolo:2019eac,Bavarsad:2009hm}. 

Previous tests of the minimal SME focussed on one operator at a time \cite{Kostelecky:2007zz, Kostelecky:2008ts}, and neglected the possible interplay between them, which instead might affect theoretical predictions and then observational constraints. In this work, we exploit the large amount of information stored in the CMB polarization spectra~\cite{planck2016-l01,BICEP:2021xfz,ACT:2020gnv} to perform a more complex analysis, accounting for different operators at the same time. In doing so, we employ the formalism recently developed by some of the authors of this work~\cite{Lembo:2020ufn}. This novel formalism allows to describe in all generality the effects of anomalous propagation of polarized radiation in terms of an effective susceptibility tensor. Any model implying anomalous propagation of radiation can be mapped into the components of this effective susceptibility tensor and the implications for the CMB power spectra can be readily derived. 

This work is timely, since it provides updated constraints on Lorentz violating coefficients using a novel mathematical formalism and state-of-the-art CMB data. Moreover, it paves the way to analogous tests with upcoming CMB data. Indeed, CMB polarization is the main observational target of next-generation CMB experiments \citep{ACT-Duivenvoorden,SPT-Guidi,BK-Karkare,SimonsObservatory:2018koc,LiteBIRD:2022cnt,CMB-S4:2016ple}.

The paper is structured as follows. In Section \ref{sec:model} we map the coefficients of the minimal SME operators which describe Lorentz violation in the radiation sector onto the effective susceptibility tensor. We define a number of phenomenological parameters related to the SME operators. This allows us to propagate the combined effects of the SME operators to the CMB spectra. We discuss the phenomenological impact of individual operators as well as their interplay. 
We then proceed to constrain the operators using data from a number of CMB experiments, as detailed in Section \ref{sec:method}. Results are reported in Section \ref{sec:Results}. 
In Section \ref{sec:Connection}, we translate the bounds obtained on the phenomenological parameters into bounds on the actual coefficients appearing in the minimal SME Lagrangian. 
We compare our results to constraints obtained both with other CMB datasets and with other kinds of observations.

\section{Imprints of Lorentz violation on the CMB spectra} \label{sec:model}

In this section, we introduce the theoretical model that describes Lorentz violating (LV) effects in the electromagnetic sector and propagate the effects to the cosmological observables of interest, namely, the CMB spectra. 
As we mentioned in the Introduction, we treat LV effects  within the SME framework \citep{Kostelecky:2002hh}, focussing on the so-called minimal SME, which only contains renormalizable operators, with mass dimension $d\leq 4$. For the photon sector and in a general spacetime with metric $g_{\mu\nu}$ this is characterized by the action
\begin{equation}
\label{eq:full-L-FLRW}
\mathcal{S} = \int d^4 x \sqrt{-g}\left[ -\frac{1}{4} F_{\mu\nu}F^{\mu\nu} + \frac{1}{2}\varepsilon^{\alpha\beta\mu\nu}A_\beta (k_{AF})_\alpha F_{\mu\nu} - \frac{1}{4} (k_F)^{\alpha\beta\mu\nu} F_{\alpha\beta}F_{\mu\nu}\right] \, ,
\end{equation}

where  we set  $\varepsilon^{\alpha\beta\mu\nu} = \epsilon^{\alpha\beta\mu\nu}/\sqrt{-g}$, with $\epsilon_{\alpha\beta\mu\nu}$ being the completely antisymmetric Levi-Civita symbol and  $g = \det(g_{\mu\nu})$. $ F_{\mu\nu}$ and $A_{\mu}$ are the field-strength tensor and the electromagnetic 4-potential, respectively. The first term in Eq.~\eqref{eq:full-L-FLRW} is just the standard Maxwell Lagrangian.
The couplings $k_{AF}$ account for operators with mass-dimension $d=3$ which violate CPT symmetries besides Lorentz symmetries. The vector $(k_{AF})_{\alpha}$ has dimensions of a mass and 4 independent components. The couplings $k_F$ govern operators with mass-dimension $d=4$ that are invariant under CPT. 
The tensor $(k_F)^{\mu\alpha\beta\gamma}$ is dimensionless and 
obeys the following symmetries
\begin{align}
(k_F)^{\mu\alpha\beta\gamma} &= -(k_F)^{\alpha\mu\beta\gamma} = -(k_F)^{\mu\alpha\gamma\beta} \, , \\
(k_F)^{\mu\alpha\beta\gamma} &= (k_F)^{\beta\gamma\mu\alpha} \, ,
\end{align}
plus a vanishing double trace, thus implying a total of 19 independent components.

Applying the Euler-Lagrange equations to the action in Eq.~\eqref{eq:full-L-FLRW} leads to the following modified Maxwell's equations:
\begin{equation}
\label{eq:modified-Maxwell}
\partial_\nu (\sqrt{-g}F^{\mu\nu}) + \varepsilon^{\mu\nu\rho\sigma}(k_{AF})_\nu \sqrt{-g} F_{\rho\sigma} + \partial_\nu \left[(k_F)^{\mu\nu\rho\sigma} \sqrt{-g} F_{\rho\sigma}\right] = 0 \, . 
\end{equation}

The usual Maxwell's theory is invariant under conformal transformations of the metric $g_{\mu\nu}\to a g_{\mu\nu}$. 
This guarantees that Maxwell's equations in a Friedmann-Lemaitre-Robertson-Walker (FLRW) Universe, described by the metric 
\begin{equation}
    g_{\mu\nu} = a^2(\tau) \eta_{\mu\nu} = a^2(\tau) \left[-d\tau^2 + d\mathbf{x}^2\right] \, ,
\end{equation}
are the same as in Minkowski spacetime with the metric $\eta_{\mu\nu}$. Here $\tau$ represents the conformal time.
In order for this invariance to be preserved by the LV theory in Eq.~\eqref{eq:full-L-FLRW}, the coefficients $(k_F)^{\alpha\beta\mu\nu}$ must transform according to \cite{Kahniashvili:2008va}:
\begin{equation}
(k_F)^{\alpha\beta\mu\nu} \rightarrow a^{-4} (k_F)^{\alpha\beta\mu\nu} \, ,
\end{equation}
such that the $a^{-4}$ factor cancels out the $a^4$ coming from $\sqrt{-g}$. Instead, the vector $(k_{AF})_\alpha$ must be invariant under the conformal transformation, since the scaling of $\sqrt{-g}$ is canceled by that of the Levi-Civita tensor $\varepsilon^{\alpha\beta\mu\nu} \propto 1/\sqrt{-g}$.

A well-known analogy exists between LV electrodynamics in vacuum and the standard Maxwell electrodynamics in an anisotropic medium, as first explored in Ref.~\cite{Carroll:1989vb} (see also Refs.~\cite{Kostelecky:2002hh, Gubitosi:2010dj}). This analogy can be exploited to define an effective susceptibility tensor $\chi_{ij}$ from the modified Ampère-Maxwell equation, i.e. the space component ($\nu = i$) of Eq.~\eqref{eq:modified-Maxwell}.\footnote{The standard Ampère-Maxwell equation in an anisotropic medium with no external sources can be written in Fourier space as \citep{fowles1989introduction}
\begin{equation}
\frac{\omega^2}{c^2} A_i + \left[\mathbf{k}\times(\mathbf{k}\times\mathbf{A})\right]_i = -\frac{\omega^2}{c^2} \chi_{ij}A^j \, .
\end{equation}
}
This reads 
\begin{align}
\label{eq:sucept}
\nonumber
\chi_{ij} = &-2(k_F)_{i0j0} - 2i\frac{c}{\omega}\epsilon_{ikj}(k_{AF})^k + 2\frac{c}{\omega} \left[-i\frac{c}{\omega}(k_{AF})_0\epsilon_{ikj} + (k_F)_{ik0j} + (k_F)_{i0kj} \right] k^k \\
& + 2\frac{c^2}{\omega^2} (k_F)_{ilkj} k^l k^k \, ,
\end{align}
where $\omega$ and $k$ are the \textit{comoving} angular frequency and wave-number, respectively:
\begin{equation}
\omega = a \omega_{phys} \, , \quad k = a k_{phys} \, .
\end{equation}
Note that the CPT-odd operator introduces in $\chi_{ij}$ only terms that are zero- and first-order in the wave-vector, whereas the CPT-even operator produces also a contribution that is quadratic in $k$. This does not come as a surprise, since it is not possible to construct a quadratic term in the wave-vector by contracting its components with those of the 3D Levi-Civita tensor and the three-vector $\mathbf{k}_{AF}$.

To evaluate the effects of the LV operators on the CMB power spectra, we first need to link the susceptibility tensor to the components of the mixing matrix in the radiative transfer equation for the Stokes parameters $Q$, $U$ and $V$ of the polarized CMB radiation. Employing the formalism developed in Ref.~\cite{Lembo:2020ufn}, the components of the susceptibility tensor can be then recast in terms of three quantities, $\rho_Q$, $\rho_U$ and $\rho_V$, describing a general mixing between the $U$ and $V$, $Q$ and $V$, $U$ and $Q$ Stokes parameters, respectively (see Eq.~(1) and Eq.~(12) of Ref.~\cite{Lembo:2020ufn}).
Using the same conventions as in Ref.~\cite{Lembo:2020ufn}, we introduce
\begin{equation}
\bar \rho_{\pm 2/V} (\tau) =(\tau - \tau_\mathrm{LS})^{-1} \int^\tau_{\tau_\mathrm{LS}} \mathrm{d}\tau^\prime \, \rho_{\pm 2/V}(\tau^\prime) \,,
\end{equation}
where $\rho_{\pm 2} = (\rho_Q \pm i\,\rho_U)/\sqrt{2}$ and $\tau_{\rm LS}$ is the conformal time at the last scattering surface. As usual in CMB analysis, we expand $(\tau_0 - \tau_\mathrm{LS}) \bar \rho_{\pm 2/V}$ in spherical harmonics\footnote{$\tau_0$ is the conformal time today.}, with expansion coefficients $b_{\pm 2/V,\,\ell m}$. Note that the $b_{V,\ell m}$ are only non-vanishing for $\ell = \{0,1\}$, whereas the $b_{\pm2,\ell m}$ are non-vanishing only for $\ell = 2$.

At this stage, it is useful to combine the expansion coefficients  $b_{\pm 2/V,\,\ell m}$ to define the following dimensionless parameters:
\begin{align}
4\pi\,\beta^2_{AF,T} &= {b_{V,00}^2}\quad  \, \text{and} \, \quad 4\pi\,\beta^2_{AF,S}  = \sum_{m} {|b_{V,1m}|^2} \,,\\
4\pi\,\beta^2_{F,E}  &= \sum_{m} {|b_{-2,2 m} + b_{2,2 m}|^2} \,,\\
4\pi\,\beta^2_{F,B}  &= \sum_{m} {|b_{-2,2 m} - b_{2,2 m}|^2} \,.
\end{align}
In fact, these parameters are directly connected to the CMB power spectra, as we will show below, and can be related to the physical parameters appearing in the action in Eq.~\eqref{eq:full-L-FLRW} as follows: 

\begin{align}
    \label{eq:beta_0}
    \beta^2_{AF,T} & = 16 c^2 \left[\bkafzero\right]^2 \, , \\
    \label{eq:beta_123}
    \beta^2_{AF,S} & = 
    \frac{16}{3}c^2 |\mathbf{\bar{k}_{AF}}|^2 = \frac{16}{3}c^2 \left(\left[(\bar{k}_{AF})_1\right]^2 + \left[(\bar{k}_{AF})_2\right]^2 + \left[(\bar{k}_{AF})_3\right]^2 \right) \, , \\
    \label{eq:beta_FE}
    \nonumber
    \beta^2_{F,E} & = \frac{64}{5} 
    \bigg[\Big((\bar{k}_F)_{3020}+(\bar{k}_F)_{3121}\Big)^2+\Big((\bar{k}_F)_{3010}-(\bar{k}_F)_{3221}\Big)^2 +\Big((\bar{k}_F)_{2010}+(\bar{k}_F)_{3231}\Big)^2\bigg] \\ 
    & \equiv \frac{64}{5} \bar{k}_{F,E}^2 \, , \\
    \nonumber
    \label{eq:beta_FB}
    \beta^2_{F,B} & = \frac{32}{15}  \bigg\{2 \Big(2 (\bar{k}_F)_{3021}+(\bar{k}_F)_{3120}-(\bar{k}_F)_{3210}\Big)^2+6\Big[\Big((\bar{k}_F)_{3120}+(\bar{k}_F)_{3210}\Big)^2 \\ 
    \nonumber
    & \quad\, + \Big((\bar{k}_F)_{3110}-(\bar{k}_F)_{3220}\Big)^2+\Big((\bar{k}_F)_{2120}-(\bar{k}_F)_{3130}\Big)^2 \\ 
    \nonumber
    & \quad\, +\Big((\bar{k}_F)_{2110}+(\bar{k}_F)_{3230}\Big)^2\Big]\bigg\} \\
    & \equiv \frac{32}{15} \bar{k}_{F,B}^2 \, ,
\end{align}
where the bar denotes quantities averaged along the line of sight, e.g. $\bkafzero \equiv \int_{\tau_{\mathrm{LS}}}^{\tau_0}  \kafzero\, d\tau$ and\footnote{As stated above, $k_F$ is dimensionless while $k_{AF}$ has the dimension of an energy in natural units. This explains the appearance of the $\omega$ factor in the expression for $(\bar{k}_F)_{ijkl}$.} $(\bar{k}_F)_{ijkl} \equiv \int_{\tau_{\mathrm{LS}}}^{\tau_0} \omega (k_F)_{ijkl} \, d\tau$. 
To derive these relations, we have assumed that the standard dispersion relation for photons holds true, i.e. $\omega = ck$. In principle, one should take into account the corrections to the dispersion relation, which are of the kind $\omega = ck\left[ 1+\mathcal{O}(k_F, k_{AF}) \right]$.
When included in Eqs.~\eqref{eq:beta_0}-\eqref{eq:beta_FB}, these corrections lead to higher-order contributions in $k_F$ and $k_{AF}$. Since LV effects are constrained to be very small \cite{Kostelecky:2008ts}, we can work at leading order in the LV coefficients, so that we can take $\omega \simeq c k$.

The phenomenological parameters of Eqs.~\eqref{eq:beta_0}-\eqref{eq:beta_FB} relate the observed CMB spectra $C^{XX}_\ell$ to those expected if no LV effects are in place, which we denote $\tilde C^{XX}_\ell$. Keeping terms up to second order in the $\beta$'s, which corresponds to working at second order in the parameters appearing in the action \eqref{eq:full-L-FLRW}, we find: 
\begin{align}
    \label{eq:clTE}
    C^{TE}_\ell &= \left(1- \frac{\calZ}{2} \right) \wt{C}^{TE}_{\ell}  \,, \\
    \label{eq:EE}
    C^{EE}_\ell &= \left(1-\calZ\right) \wt{C}^{EE}_{\ell} + \sum_{\ell_1} \, \calK^{11}_{\ell_1\ell}  \wt{C}^{EE}_{\ell_1} + \sum_{\ell_1} \, \calK^{22}_{\ell_1\ell}  \wt{C}^{BB}_{\ell_1}  \,, \\
    \label{eq:BB}
    C^{BB}_\ell &= \left(1-\calZ\right) \wt{C}^{BB}_{\ell} + \sum_{\ell_1} \,\calK^{11}_{\ell_1\ell}  \wt{C}^{BB}_{\ell_1} + \sum_{\ell_1} \, \calK^{22}_{\ell_1\ell}  \wt{C}^{EE}_{\ell_1} \,,  \\
    \label{eq:EB}
    C^{EB}_\ell &= \sqrt{\beta^2_{AF,T}} \left(\wt{C}^{EE}_{\ell}-\wt{C}^{BB}_{\ell}\right)  \, , \\[0.3cm]
    \label{eq:TB}
    C^{TB}_\ell &= \sqrt{\beta^2_{AF,T}} \wt{C}^{TE}_{\ell}  \, , \\[0.3cm]
    \label{eq:clVV}
    C_\ell^{VV} &= \sum_{\ell_1} \,\calK_{\ell_1 \ell}^{33} \tilde{C}_{\ell_1}^{EE} + \sum_{\ell_1} \,\calK_{\ell_1 \ell}^{44} \tilde{C}_{\ell_1}^{BB} \, , \\
     C_\ell^{EV} &= C_\ell^{BV} = 0 \, ,    \label{eq:clEV}
\end{align}
where 
\begin{align}
    \label{eq:Z}
    \calZ &= \beta^2_{AF,T}+\beta^2_{AF,S} + \frac{ \left(\beta^2_{F,E}+\beta^2_{F,B} \right)}{4} \, , \\
    \label{eq:K11}
    \sum_{\ell_1} \, \calK^{11}_{\ell_1\ell}  \wt{C}^{XX}_{\ell_1} &= \beta^2_{AF,S}\, \frac{4}{\ell+\ell^2} \wt{C}^{XX}_{\ell} \, , \\
    \label{eq:K22}
	\sum_{\ell_1} \, \calK^{22}_{\ell_1\ell}  \wt{C}^{XX}_{\ell_1} &= \beta^2_{AF,T}\, \wt{C}^{XX}_{\ell} +  \beta^2_{AF,S}\, \left( \frac{\ell^2-4}{\ell(2\ell+1)} \wt{C}^{XX}_{\ell-1}+
	\frac{(\ell-1)(\ell+3)}{(\ell+1)(2\ell+1)} \wt{C}^{XX}_{\ell+1} \right) \, , \\
	\nonumber
	\label{eq:K33}
	\sum_{\ell_1} \, \calK^{33(44)}_{\ell_1\ell}  \wt{C}^{XX}_{\ell_1} &= \beta^2_{F,B(E)} \left(\frac{(\ell-2)(\ell-3)}{4 (4\ell^2-1)}\tilde{C}_{\ell-2}^{XX} + \frac{3(\ell^2+\ell-2)}{2(4\ell^2+4\ell-3)}\tilde{C}_{\ell}^{XX} \right. \\
	& \left. \quad + \frac{(\ell+3)(\ell+4)}{4(4\ell^2+8\ell+3)}\tilde{C}_{\ell+2}^{XX} \right) + \beta^2_{F,E(B)} \left(\frac{\ell-2}{2(2\ell+1)}\tilde{C}_{\ell-1}^{XX} + \frac{3+\ell}{2(2\ell+1)}\tilde{C}_{\ell+1}^{XX} \right) \, .
\end{align}
By inspecting Eqs.~\eqref{eq:clTE}-\eqref{eq:clEV} we can identify the effects of different classes of LV operators:
\begin{itemize}
    \item the CPT-odd operators, parametrized by $\beta^2_{AF,T}$ and $\beta^2_{AF,S}$, lead to the well-known cosmic birefringence effect. In particular, $\beta^2_{AF,T}$, related to the time component of the 4-vector $k_{AF}$, gives rise to isotropic birefringence~\citep{Lue:1998mq,Feng:2006dp,Kostelecky:2008be,Gubitosi:2009eu,Pagano:2009kj,Gruppuso:2015xza,Minami:2020odp}, which produces non-vanishing EB and TB spectra and the mixing between EE and BB spectra. 
    Anisotropic birefringence~\citep{Gluscevic:2012me,Gubitosi:2011ue,Contreras:2017sgi,SPT:2020cxx,Gruppuso:2020kfy,Bortolami:2022whx} is induced by the parameter $\beta^2_{AF,S}$, related to the space components of $k_{AF}$. 
    This mixes the EE and BB spectra by  introducing a coupling among different multipoles (i.e. off-diagonal correlations), such that the $\ell$-th multipole is coupled to both the $(\ell-1)$-th and $(\ell+1)$-th ones;
    \item the VV spectrum is  sourced from EE and BB spectra when the CPT-even operators are present.
    Similarly to what observed for anisotropic birefringence, a coupling between different multipoles is induced. In this case, it affects all the multipoles between the $(\ell-2)$-th and the $(\ell+2)$-th. 
    Note that the VV spectrum is the only one which, if measured, could break the degeneracy between $\beta^2_{F,E}$ and $\beta^2_{F,B}$, since in the other spectra only the sum of these two parameters comes into play. 
    In this model, no mixing is predicted between V modes and E- or B-modes;
    \item both the CPT-even and CPT-odd operators rescale the EE, BB and TE spectra via the parameter $\mathcal{Z}$. 
\end{itemize}

The modifications to the CPT-even linear polarization spectra, Eqs.~\eqref{eq:clTE}-\eqref{eq:BB}, and the introduction of the circular polarization spectrum, Eq.~\eqref{eq:clVV}, have been implemented in a customized version of the Boltzmann code CAMB \citep{Lewis:1999bs,Howlett:2012mh}, hereafter \cambcpt\footnote{We make the code publicly available at this link: \url{https://github.com/sgiardie/CAMB_CPT}}. 
In the code, we have treated gravitational lensing of the CMB and the modifications induced by the extra terms in the action,  Eq.~\eqref{eq:full-L-FLRW}, as two distinct effects. In principle, these two mechanisms should be propagated simultaneously along the line of sight, see for example Ref.~\citep{Gubitosi:2014cua}. However, the kernel of the lensing effect is peaked at low redshift while the effect of the LV electrodynamics on the CMB is integrated from the last scattering surface and acts as a small correction. Therefore, as far as B-modes are concerned, we can safely rotate the tensor signal and then add the B-mode lensing contribution computed assuming no rotation.
Instead, regarding EE and TE, we apply corrections to the lensed spectra. This is justified by considering that, for the noise level of current CMB experiments and even for the noise level of SO and LiteBIRD\footnote{For deeper surveys, such as CMB-S4, these approximations should be reconsidered.}, there is a negligible difference between modifying the lensed spectra (i.e., applying the Lorentz-violating effect \textit{after} the lensing contribution is included) and acting on the unlensed ones \textit{before} adding lensing, see~\cite{Lembo:2020ufn,Gubitosi:2014cua}.

\begin{figure}
\centering
\includegraphics[width=0.8\textwidth]{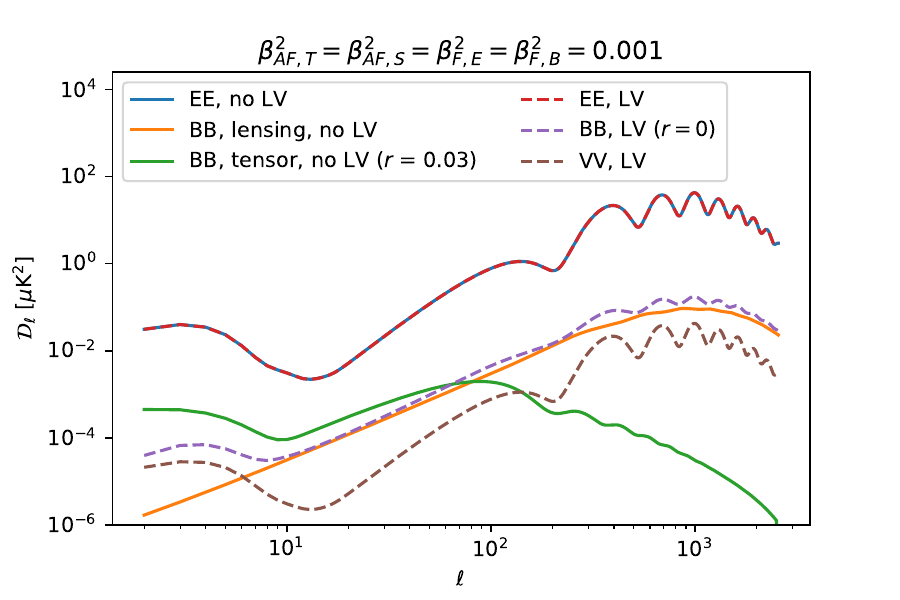}
\caption{Standard CMB power spectra in solid lines (no LV), with Lorentz violating effects (LV) in dashed lines. The LV spectra are generated according to Eqs.~\eqref{eq:clTE}-\eqref{eq:clVV} with $\beta^2_{AF,T} = \beta^2_{AF,S} = \beta^2_{F,E} = \beta^2_{F,B} = 0.001$ using \cambcpt. The VV spectrum is non-vanishing only in the LV case, sourced by both the standard EE and BB spectra. Note that the EE spectra are almost overlapped and  practically indistinguishable 
with this choice of the LV parameters.} \label{fig:spectra}
\end{figure}

Figure \ref{fig:spectra} shows a comparison between the standard CMB spectra (solid) and those obtained with \cambcpt\ by setting  all the $\beta^2$ parameters equal to 0.001 (dashed). The most relevant feature is the leakage of E- into B-modes. Another clear effect is the VV power spectrum mostly sourced by the E modes. The linear-polarization spectra are also rescaled by the $\mathcal{Z}$ factor in Eq.~\eqref{eq:Z}, which depends on all the $\beta^2$ parameters. The latter effect is barely visible on the scale of the figure.

\section{Analysis method and dataset} \label{sec:method}

We perform a Monte Carlo Markov Chain (MCMC) analysis to obtain constraints on the Lorentz-violating parameters $\beta^2_{AF,T}, \beta^2_{AF,S}, \beta^2_{F,E}, \beta^2_{F,B}$ jointly with other cosmological, foreground and nuisance parameters. To this scope, the code \cambcpt\ has been interfaced with the MCMC sampler \Cobaya\ \cite{Torrado:2020dgo}. Using the Gelman-Rubin convergence statistics \cite{Gelman:1992zz}, we have assumed that our MCMC chains have reached convergence when $R-1 \sim 0.01$.

We analyze the following data:

\begin{itemize}
    \item \Planck\ 2018: \Planck\ temperature and polarization power spectra \cite{planck2016-l05}, and lensing reconstruction power spectrum \cite{planck2016-l08}, from the \Planck\ 2018 legacy release.
    \item BICEP/Keck 2018 (BK18): combination of all the B modes data collected by BICEP2, Keck Array and BICEP3 experiments until the 2018 season \cite{BICEP:2021xfz}.
    \item ACT: Atacama Cosmology Telescope temperature and polarization power spectra as published in the Data Release 4 \citep{ACT:2020gnv}. Since the ACT data are always used in combination with \Planck, following the prescription of the ACT collaboration, we only consider multipoles larger than $1800$ in temperature. For more details see section 6.2.3 of \citep{ACT:2020gnv}.
    \item VV: V modes power spectra as published by CLASS \citep{Padilla:2019dhz} and SPIDER \citep{SPIDER:2017kwu} experiments.
\end{itemize}

For \Planck, BICEP/Keck and ACT we employ the official likelihood packages released by the respective collaborations \citep{planck2016-l05,planck2016-l08,BICEP:2021xfz,ACT:2020gnv}. For the V-modes data, a simple custom-made likelihood has been added to the framework. The $\chi^2$ for the V modes is computed as:
\begin{equation} \label{eq:chi2}
\chi^2_{VV} = \sum_b \frac{(D^{VV}_{b, \rm{theory}}-D^{VV}_{b, \rm{data}})^2}{\sigma^2_b} \, ,
\end{equation}
where $D^{VV}_{b, \rm{data}}$ and $D^{VV}_{b, \rm{theory}}$ are the data and the binned theory respectively and $\sigma^2_b$ is the error on the bandpowers.\footnote{The theoretical power spectra are binned with flat window function in $\ell(\ell+1)/2\pi$.}
Since CLASS and SPIDER are both completely noise dominated, we can safely add together their respective $\chi^2$ computed as in Eq.~\eqref{eq:chi2}.

In our analysis we consider the following data combinations:\footnote{Notice that we do not consider the combination \Planck\ 2018 + BK18 + ACT + SPIDER + CLASS since the inclusion of V-modes data does not add any constraining power, see the discussion in Section \ref{sec:Results} for more details.}
\begin{enumerate}[label=(\roman*)]
    \item \Planck\ 2018;
    \item \Planck\ 2018 + BK18;
    \item \Planck\ 2018 + BK18 + CLASS + SPIDER; 
    \item \Planck\ 2018 + BK18 + ACT.
\end{enumerate}

The $\Lambda$CDM+$r$ model (i.e., allowing for non-vanishing primordial gravitational waves with amplitude set by the tensor-to-scalar ratio $r$) provides our baseline scenario, unless otherwise stated. See Ref.~\cite{planck2016-l06} for details about parametrization, theoretical assumptions and priors used. 
For the foreground and nuisance parameters, we follow the prescriptions provided by \Planck\ \cite{planck2016-l06} and BICEP \cite{BICEP:2021xfz} collaborations. In addition to the baseline, we consider the $\beta^2$ parameters defined in Eqs.~\eqref{eq:beta_0}-\eqref{eq:beta_FB}. On those parameters we impose uniform positive priors. Further model extensions are not considered in this work.

\section{Constraints on phenomenological parameters}
\label{sec:Results}

In this section, we present the constraints derived on $\beta^2_{AF,T}$, 
$\beta^2_{AF,S}$,  $\beta^2_{F,E}$ and $\beta^2_{F,B}$ using the aforementioned datasets and parametrizations. 

\subsection{Constraints on CPT-odd terms only}
\begin{figure}
\centering
\includegraphics[width=0.8\textwidth]{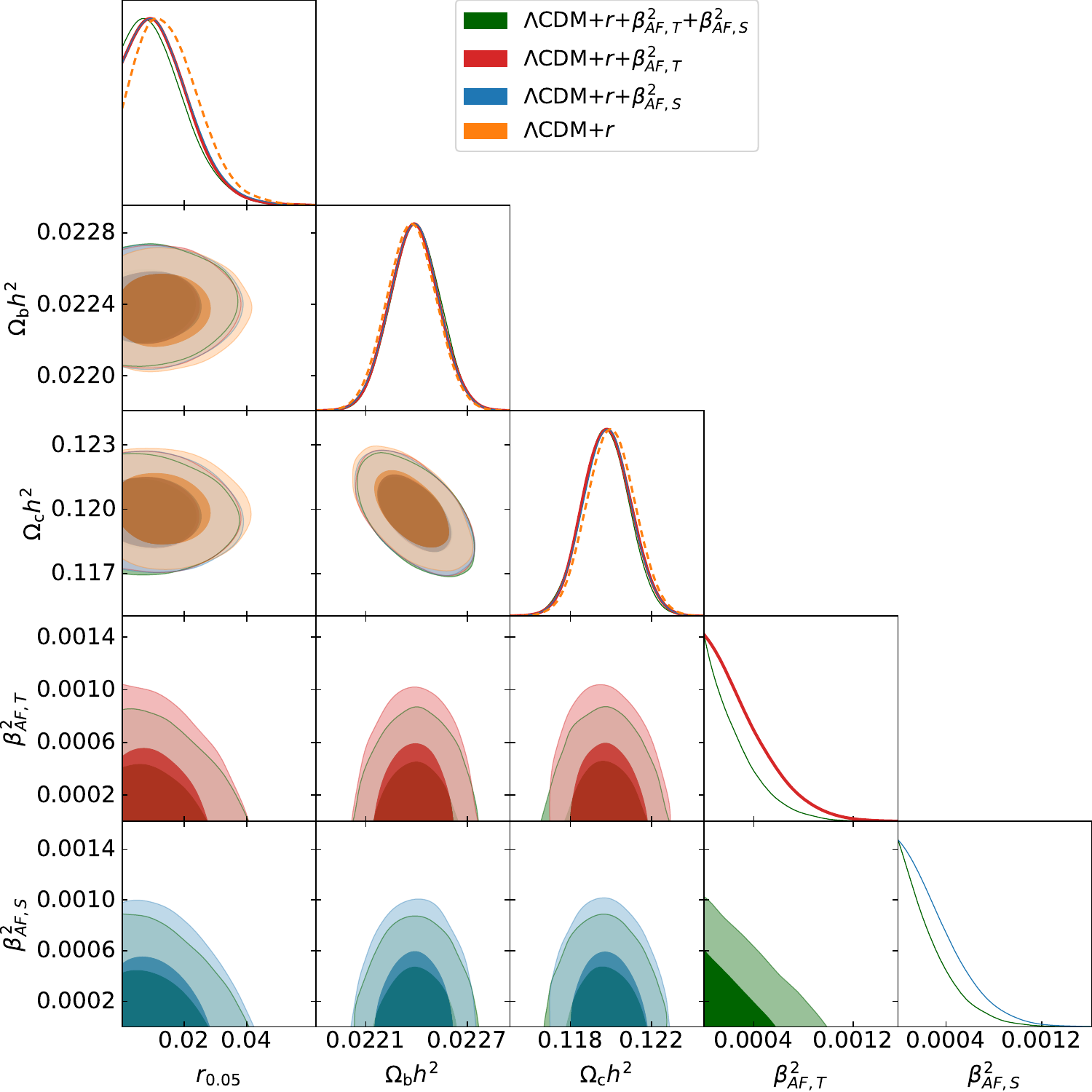}
\caption{One and two-dimensional posterior probability distributions for a subset of parameters varied in the MCMC analysis. We report the constraints obtained when assuming a $\Lambda$CDM+$r$+$\beta^2_{AF,T}$+$\beta^2_{AF,S}$ model (in green), $\Lambda$CDM+$r$+$\beta^2_{AF,T}$ (in red), $\Lambda$CDM+$r$+$\beta^2_{AF,S}$ (in blue)  and $\Lambda$CDM+$r$ (in orange) using \Planck\ TTTEEE+lensing+BK18 data. Note the tighter limit on $r$ when one of the $\beta^2_{AF}$ parameters is allowed to vary with respect to the case in which they are both equal to zero. Opening to both $\beta^2_{AF}$ further improves the individual constraints on $\beta^2_{AF,T}$, $\beta^2_{AF,S}$ and $r$, see the main text for a detailed discussion.} \label{fig:lcdm+r+beta_AF}
\end{figure}

As a first step in our analysis, we consider only the CPT-odd term in Eq.~\eqref{eq:full-L-FLRW} and fix to zero the parameters related to the CPT-even term. The effect of this term on the CMB spectra is encoded in two parameters $\beta^2_{AF,T}$ and $\beta^2_{AF,S}$ and leads to isotropic and anisotropic birefringence effects, respectively. 
In Figure \ref{fig:lcdm+r+beta_AF}, we show the two-dimensional and one-dimensional posterior probability distributions of a subset of cosmological parameters, including $\beta^2_{AF,T}$ and $\beta^2_{AF,S}$, explored in the analysis with the combination of \Planck + BK18 data. The baseline model is given by the $\Lambda$CDM+$r$ cosmology. To better elucidate the effect of $\beta^2_{AF,T}$ and $\beta^2_{AF,S}$ on the constraints of the remaining parameters, we also vary them one at the time while fixing the other to zero. 
We note that varying either $\beta^2_{AF,T}$ or $\beta^2_{AF,S}$ has equivalent impact on the constraints on other cosmological parameters. This is due to the fact that both $\beta^2_{AF,T}$ and $\beta^2_{AF,S}$ lead to qualitatively equivalent modifications of the BB spectrum. Indeed, an inspection of Eq.~\eqref{eq:BB} and Eq.~\eqref{eq:K22} shows that the overall effect produced by non-vanishing $\beta^2_{AF,T}$ or $\beta^2_{AF,S}$ is an effective rotation of E-modes into B-modes. Such rotation competes with $r$ in increasing the power in B-modes (see Figure \ref{fig:spectra}, where the two $\beta^2_{AF}$ and $r$ enhance the reionization and recombination bumps in the BB power spectrum). This explains why the marginalization over $\beta^2_{AF,T}$ and $\beta^2_{AF,S}$ tightens the constraints on $r$ with respect to those obtained in the $\Lambda$CDM+$r$ baseline analysis. 

\begin{figure}
\centering
\subfloat[]{\includegraphics[width=0.57\textwidth]{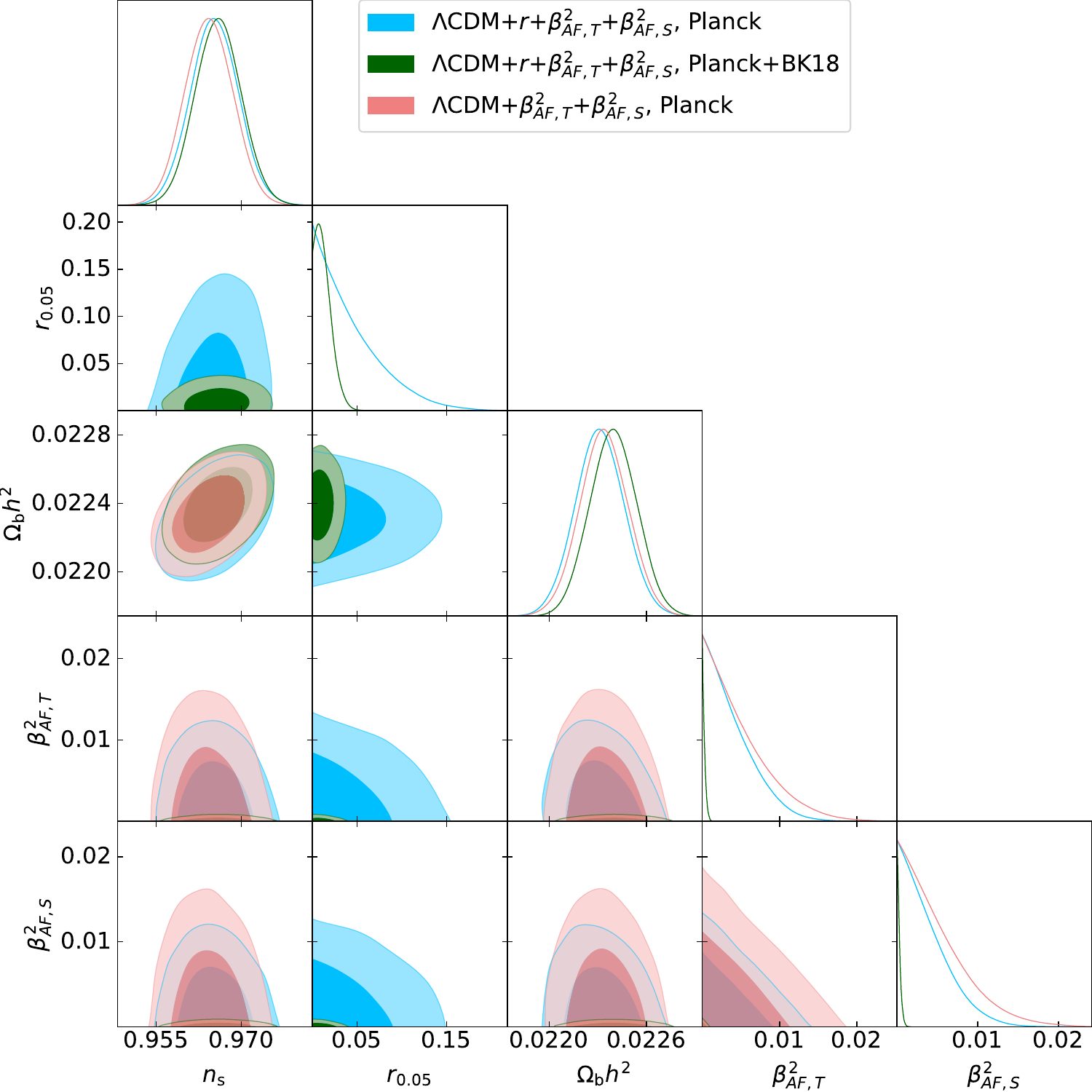}\label{fig:lcdm+beta_AF_Plonly_vs_PlBK18}} \quad
\subfloat[]{\includegraphics[width=0.4\textwidth]{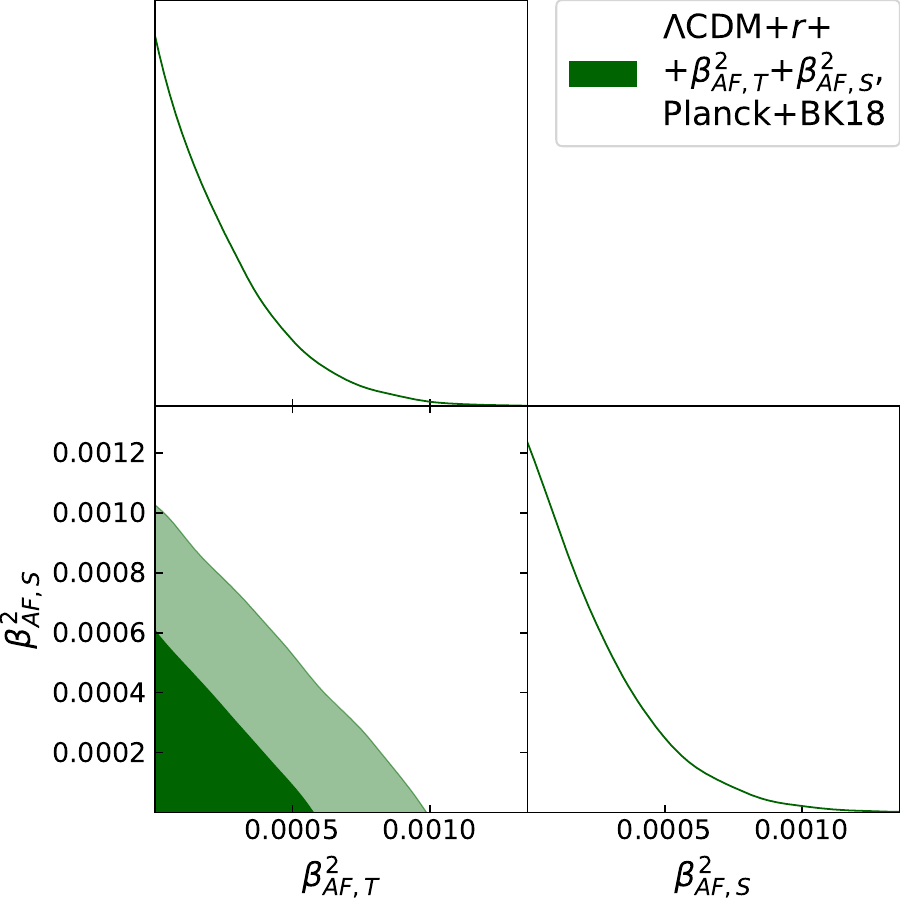}\label{fig:lcdm+beta_AF_PlBK18_only}} \quad
\caption{On the left, one and two-dimensional posterior probability distributions for a subset of parameters varied in the MCMC analysis. We report the constraints obtained when assuming a $\Lambda$CDM+$\beta^2_{AF,T}+\beta^2_{AF,S}$ (in pink) and  a $\Lambda$CDM+$r$+$\beta^2_{AF,T}+\beta^2_{AF,S}$ (in cyan and green) models. The former using only \Planck\ TTTEEE+lensing dataset, while the latter using both \Planck\ TTTEEE+lensing and \Planck\ TTTEEE+lensing+BK18 datasets. Note how much the constraints on the $\beta^2_{AF}$ parameters improve when we include BK18 data. On the right, a zoom-in showing the constraints on $\beta^2_{AF,T}$ and $\beta^2_{AF,S}$ using \Planck\ TTTEEE+lensing+BK18 datasets.} \label{fig:lcdm+beta_AF_Plonly_vs_PlBK18_tot}
\end{figure}

Even though in Figure \ref{fig:lcdm+r+beta_AF} we report results from \Planck +BK18, the two $\beta^2_{AF}$ could be also constrained with \Planck\ data only, exploiting their effect on E-mode polarization. However, the resulting bounds on $\beta^2_{AF,T}$ and $\beta^2_{AF,S}$ are nearly an order-of-magnitude broader than those obtained when adding BK18 to \Planck\ data. This is due to the lack of constraining power from B-modes which are more strongly affected by the two $\beta^2_{AF}$.
In Fig.~\ref{fig:lcdm+beta_AF_Plonly_vs_PlBK18}, we show the constraints on a subset of parameters and compare the results obtained with \Planck\ data only in $\Lambda$CDM+$\beta^2_{AF}$ and $\Lambda$CDM+$r$+$\beta^2_{AF}$ with those obtained with the combination of \Planck + BICEP/Keck data in $\Lambda$CDM+$r$+$\beta^2_{AF}$. As expected, the bounds on $\beta^2_{AF}$ are tightened when $r$ is varied jointly with the CPT-odd parameters, even if using \Planck\ data only. However, the improvement is dramatic when BICEP/Keck data are added to the analysis. In Fig.~\ref{fig:lcdm+beta_AF_PlBK18_only} we show a zoom-in of the lower right triangle of Figure \ref{fig:lcdm+beta_AF_Plonly_vs_PlBK18} to better appreciate the impact of BICEP/Keck data on the constraints on the $\beta^2_{AF}$. 
We stress again that no V-modes are sourced by the CPT-odd term of the Lagrangian.

\subsection{Constraints on CPT-even terms only}
We now focus on the CPT-even term of the action in Eq.~\eqref{eq:full-L-FLRW}. The effects on the CMB spectra are in this case encoded by the two parameters $\beta^2_{F,E}$ and $\beta^2_{F,B}$, which are responsible for an overall rescaling of the TE, EE and BB $\tilde{C}_\ell$s via the parameter $\mathcal{Z}$, see Eqs.~\eqref{eq:clTE},~\eqref{eq:EE},~\eqref{eq:BB}. 
If we restrict our analysis to consider only linear polarization, the impact of the two $\beta^2_{F,E/B}$ is degenerate.
However, the CPT-even term sources a degree of circular polarization from a mixing of E- and B-modes appropriately rescaled by $\beta^2_{F,E}$ and $\beta^2_{F,B}$, see Eq.~\eqref{eq:clVV}. The sourcing of V-modes could in principle be used to individually constrain the $\beta^2_{F,E/B}$, provided that a V-mode  experiment puts statistically significant bounds on the VV signal. However, the signal-to-noise ratio in the SPIDER and CLASS data is insufficient to put significant bounds on the two parameters. 
This is shown in Figure \ref{fig:lcdm+r+beta_F_complete} of the App.~\ref{app:plots}, where the posterior distributions on cosmological parameters, including $\beta^2_{F,E}$, with and without V-mode data are perfectly overlapping. We expect this to be exactly the same for $\beta^2_{F,B}$, since in the absence of sensitive enough V-mode data both $\beta^2_{F,E}$ and $\beta^2_{F,B}$ parameters are constrained through the rescaling of TE, EE and BB spectra within $\mathcal{Z}$. 
Therefore, in the following, we neglect the contribution of V modes data and we quote results for the effective parameter $\beta^2_F$, defined as 
\begin{equation}
    \label{eq:beta-F}
    \beta^2_F \equiv \beta^2_{F,E}+\beta^2_{F,B} \, .    
\end{equation}

\begin{figure}
\centering
\includegraphics[width=0.6\textwidth]{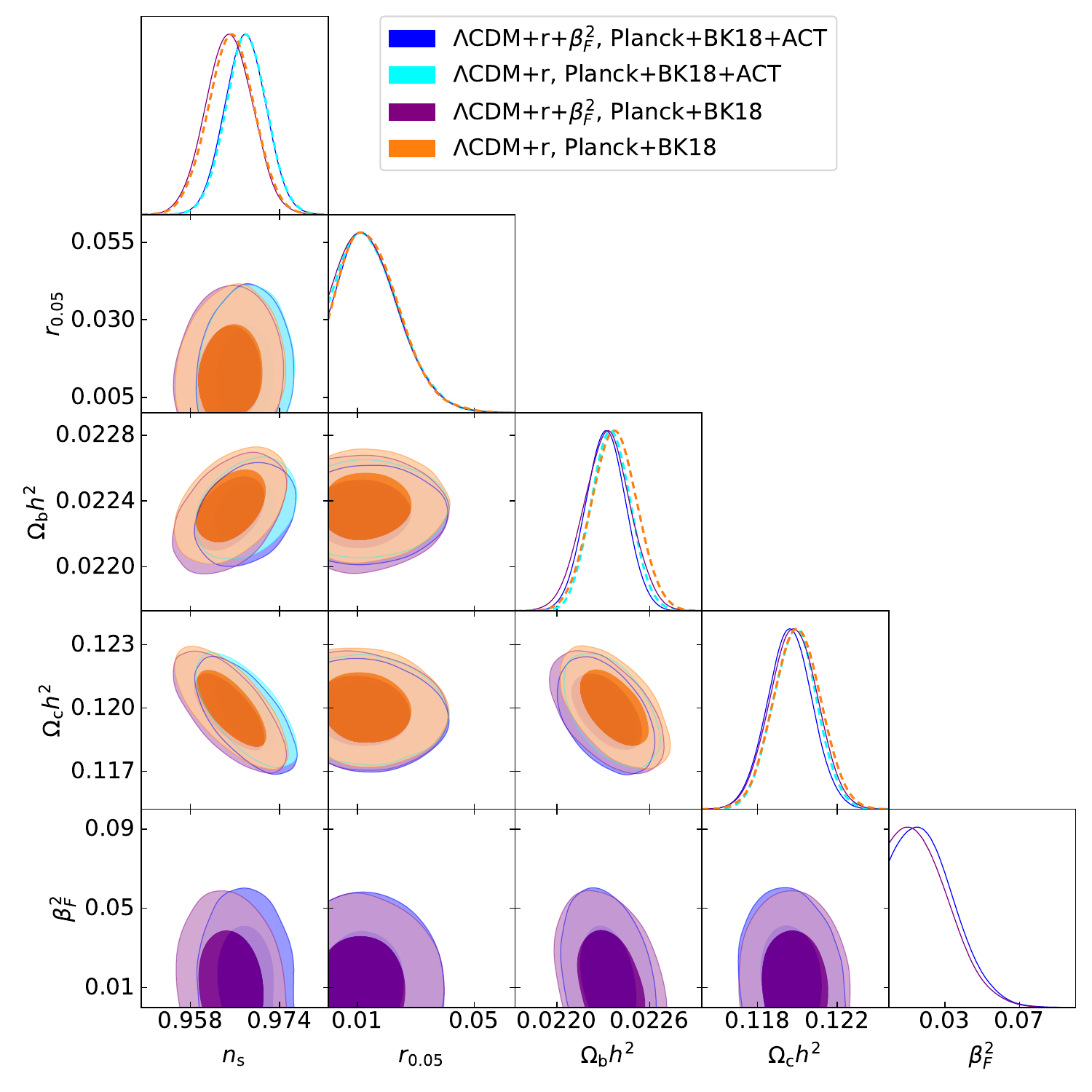}
\caption{One and two-dimensional posterior probability distributions for a subset of parameters varied in the MCMC analysis. We report the constraints obtained when assuming $\Lambda$CDM+$r$+$\beta^2_{F}$ (in purple when using the \Planck \ TTTEEE+lensing+BK18 dataset, in blue when adding ACT) and $\Lambda$CDM+$r$ (in dashed orange and dashed cyan respectively). Since not enough constraining power comes from current V-mode data, we are not able to disentangle the effects of $\beta^2_{F,E}$ and $\beta^2_{F,B}$, and we can only set a limit on their combination $\beta^2_F = \beta^2_{F,E}+\beta^2_{F,B}$. Note the shifts in the posteriors of $\Omega_b h^2$ and $\Omega_c h^2$ when considering the $\Lambda{\rm CDM}+r+\beta^2_{F}$ extension, see the main text for a detailed discussion.} \label{fig:lcdm+r+betaF}
\end{figure}

In Fig.~\ref{fig:lcdm+r+betaF}, we show 2D and 1D posterior probabilities of a subset of cosmological parameters explored with the combination of \Planck +BK18 and \Planck +BK18+ACT data. We compare the results within the $\Lambda$CDM$+r+\beta^2_F$ model and the baseline $\Lambda$CDM$+r$ model. 

 Differently from what discussed for the CPT-odd parameters, we do not see any improvement in the bounds on $r$ when $\beta^2_F$ is varied. In this case, we expect a positive correlation between $r$ and $\beta^2_F$, contrarily to what happens with the $\beta^2_{AF}$. Indeed, a non-vanishing $\beta^2_F$ reduces the amplitude of the BB spectrum, which could be compensated by  higher values of $r$. However, we do not appreciate such a correlation in Fig.~\ref{fig:lcdm+r+betaF}. The reason is that most of the constraining power on $\beta^2_F$ comes from TE and EE spectra, making any degeneracy with $r$ undetectable. Indeed, the sensitivity on $\beta^2_F$ from T- and E-modes only is at the same level as that on $\beta^2_{AF,T/S}$, being driven by the scaling in amplitude of EE and TE spectra.
In Fig.~\ref{fig:lcdm+r+betaF}, we also note a shift in $\Omega_b h^2$ and $\Omega_c h^2$ with respect to the constraints obtained when $\beta^2_F=0$. The shifts can be easily explained when considering the impact of the parameters on the shape of the TE and EE spectra. 
The main effect of the non-vanishing $\beta^2_F$ on the polarization power spectra is to rescale their overall amplitude through $\calZ$ in Eq.~\eqref{eq:Z}. 
A change in $\Omega_b h^2$, instead, affects the amplitude of the TE and EE acoustic oscillations both in the photon density field (by modifying the inertia of the baryon-photon fluid, which is relevant for the temperature transfer function) and in the photon velocity field (as a result of the change in the density), which is relevant for the E-polarization transfer function. From these considerations, we can understand the correlation between $\beta^2_F$ and $\Omega_b h^2$. 
At sub-degree scales (high multipoles $\ell$), a change in $\Omega_b h^2$ modifies the damping angular scale since a different baryon density affects the photon mean free path. As a result, the power at small scales is more or less suppressed depending on the value of $\Omega_b h^2$. This effect goes in the opposite direction of the change in the amplitude of the first peaks: a lower value of $\Omega_b h^2$ increases the amplitude of the oscillations at intermediate scales and suppresses the power at small scales. 
A similar effect at intermediate scales is provided by $\Omega_{c}h^2$. A decrease of the latter delays the onset of matter-radiation equality, thus shifting to larger scales the boosting effect due to radiation driving on the acoustic oscillations. Therefore, we expect $\Omega_{c}h^2$ to decrease when allowing for a non-vanishing $\beta^2_F$.  
The inclusion of ACT data causes the same shift of $\Omega_{b}h^2$ and $\Omega_{c}h^2$ when sampling over $\beta^2_F$, as can be seen in Fig.~\ref{fig:lcdm+r+betaF}.
Moreover, the limit on $\beta^2_F$ is broader. This is likely driven by the known preference of ACT for larger $A_s$ and $n_s$ \citep{ACT:2020gnv}, which can be compensated by a larger value of $\beta^2_F$.

\subsection{Joint constraints on CPT-odd and CPT-even terms}
\begin{figure}
\centering
\includegraphics[width=0.7\textwidth]{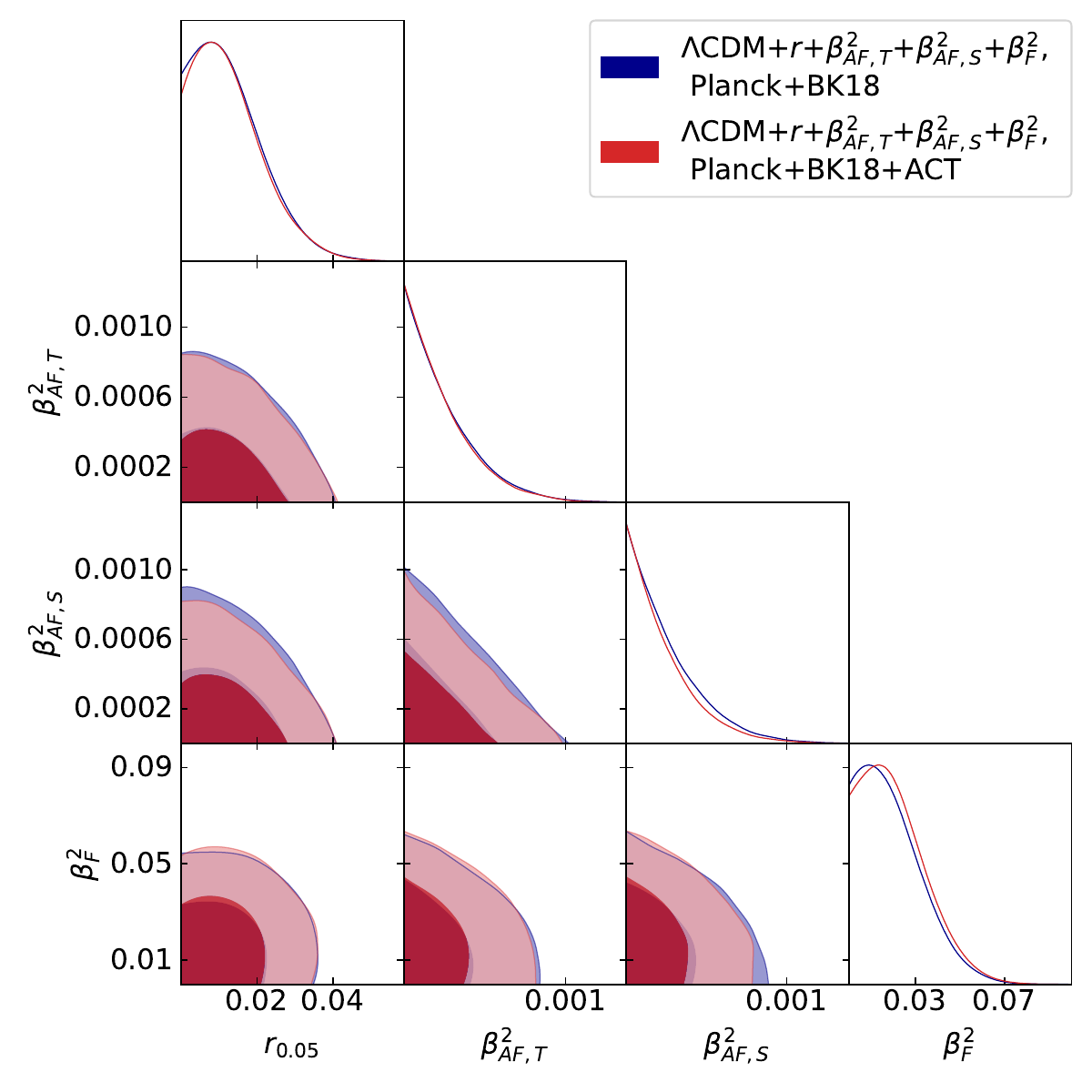}
\caption{One and two-dimensional posterior probability distributions for the LV parameters $\beta^2$ varied in the MCMC analysis. We report the constraints obtained when assuming a $\Lambda$CDM+$r$+$\beta^2_{AF,T}$+$\beta^2_{AF,S}$+$\beta^2_{F}$ model using \Planck\ TTTEEE+lensing+BK18 (in dark blue) and \Planck\ TTTEEE+lensing+BK18+ACT (in red). Including ACT data weakens the constraints mostly on $\beta^2_F$, see the main text for a detailed discussion.}
\label{fig:lcdm+r+beta_ACT}
\end{figure}

Finally, we investigate the case in which all the CPT-even and CPT-odd parameters are varied jointly. This allows us to investigate how the interplay between the effects induced by different operators affects the constraints on the LV parameters. We have collected the 95\% CL on $r$, $\beta^2_{AF,T}$, $\beta^2_{AF,S}$, $\beta^2_{F}$ for the cases analyzed in Tab. \ref{tab:beta_results}.
The posteriors on all the cosmological parameters, including those not quoted in this Section, can be found in the Appendix \ref{app:plots}. 
Figure \ref{fig:lcdm+r+betaAF+betaF} shows the 2D and 1D posterior probabilities of a subset of cosmological parameters plus the $\beta^2$s assuming a $\Lambda$CDM+$r$+$\beta^2_{AF,T}$+$\beta^2_{AF,S}$+$\beta^2_{F}$ model. For comparison, we have also included the posteriors for the $\Lambda$CDM+$r$+$\beta^2_{AF,T}$+$\beta^2_{AF,S}$ and $\Lambda$CDM+$r$+$\beta^2_{F}$ models.
On the one hand, we see that the bounds on $\beta^2_{F}$ improve when all the $\beta^2$ are allowed to vary. In fact, in absence of V-mode data, the only effect of $\beta^2_{F}$ is to contribute to the rescaling of the CMB spectra via $\calZ$, in the same way as $\beta^2_{AF,T}$ and $\beta^2_{AF,S}$ do. On the other hand, the constraints on $\beta^2_{AF,T}$ and $\beta^2_{AF,S}$ do not  improve significantly when the two parameters are varied jointly with $\beta^2_F$. In fact, besides rescaling the spectra, they also induce a mixing between E and B modes, which allows to disentangle them from $\beta^2_F$. Note again the improved bounds on $r$ when $\beta^2_{AF,T}$ and $\beta^2_{AF,S}$ are varied. The inclusion of ACT mostly affects the constraint on $\beta^2_F$ (see Figure \ref{fig:lcdm+r+beta_ACT}), as discussed before.

\begin{figure}
\centering
\includegraphics[width=0.7\textwidth]{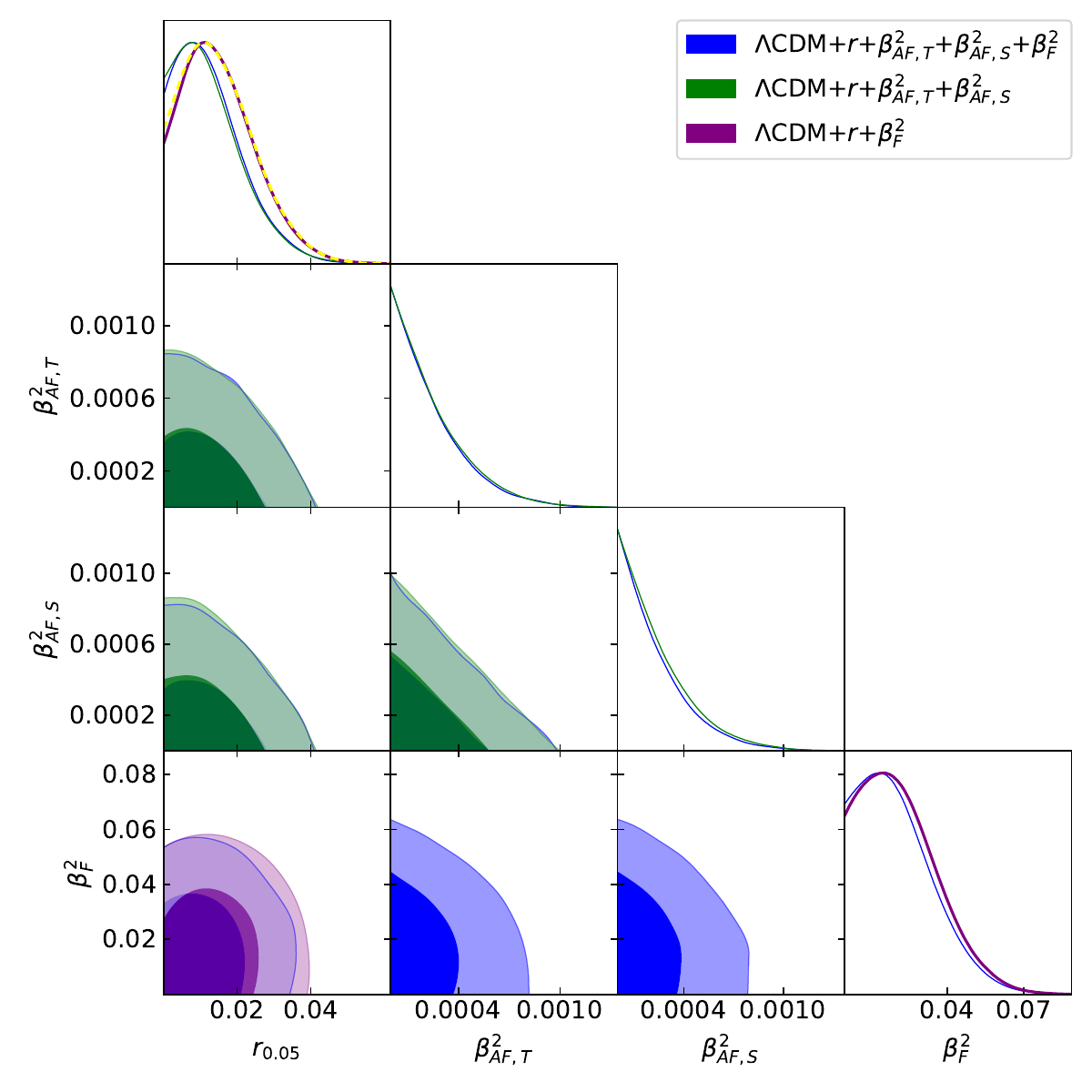}
\caption{One and two-dimensional posterior probability distributions for the LV parameters $\beta^2$ varied in the MCMC analysis. We report the constraints obtained when assuming $\Lambda$CDM+$r$+$\beta^2_{AF,T}$+$\beta^2_{AF,S}$+$\beta^2_{F}$ (in dark blue), $\Lambda$CDM+$r$+$\beta^2_{AF,T}$+$\beta^2_{AF,S}$ (in green) and $\Lambda$CDM+$r$+$\beta^2_{F}$ (in purple),
using the \Planck\ TTTEEE+lensing+BK18+ACT dataset. The posterior in dashed yellow is the reference for the $\Lambda$CDM+$r$ case using same dataset. The joint marginalization over all the $\beta^2$ parameters improves the constraints on $\beta^2_F$, while keeping unchanged those on $r$ and the $\beta^2_{AF}$ parameters.} \label{fig:lcdm+r+betaAF+betaF}
\end{figure}

\begin{table}
\centering 
\begin{tabular}{l l c c c c} 
\hline 
\rule{0pt}{11pt}
Dataset & Model (\lcdm+) & \makecell{$r\times10^{2}$ 
} &  \makecell{$\beta^2_{AF,T}$\\ $\times10^{2}$ 
} &  \makecell{$\beta^2_{AF,S}$\\ $\times10^{2}$ 
} & \makecell{$\beta^2_{F}$\\ $\times10^{2}$ 
} \\
\hline
\hline 
\rule{0pt}{11pt}
\hspace{-1.2mm}\Planck & $\beta^2_{AF,T}$+$\beta^2_{AF,S}$ & - & $< 1.29$ & $< 1.28$  & - \\
\Planck & $r$+$\beta^2_{AF,T}$+$\beta^2_{AF,S}$ &  $< 11.5$ & $< 0.987$ & $< 0.953$  & - \\
\Planck+BK18 & $r$ & $< 3.36$ & - & - & - \\
\Planck+BK18 & $r$+$\beta^2_{AF,T}$ & $< 3.07$ & $< 0.0813 $  & - & - \\
\Planck+BK18 & $r$+$\beta^2_{AF,S}$ & $< 3.13$ & -  & $< 0.0805$ & - \\
\Planck+BK18 & $r$+$\beta^2_{AF,T}$+$\beta^2_{AF,S}$ & $< 3.00$ & $< 0.0673$  & $< 0.0697$ & - \\
\Planck+BK18 & $r$+$\beta^2_{F}$ & $< 3.36$ & -  & - & $< 4.76 $ \\
\Planck+BK18 & $r$+$\beta^2_{AF,T}$+$\beta^2_{AF,S}$+$\beta^2_{F}$ & $< 3.02$ & $< 0.0675$  & $< 0.0692$ & $< 4.60$ \\
\Planck+BK18+VV & $r$+$\beta^2_{F}$ & $< 3.36$ & -  & - & $< 4.73 $ \\
\Planck+BK18+ACT & $r$+$\beta^2_{AF,T}$ & $< 3.11$ & $ < 0.0765$ & - & - \\
\Planck+BK18+ACT & $r$+$\beta^2_{AF,S}$ & $< 3.11$ & -  & $< 0.0765$ & - \\
\Planck+BK18+ACT & $r$+$\beta^2_{AF,T}$+$\beta^2_{AF,S}$ & $< 3.03$ & $< 0.0665$  & $< 0.0668$  & - \\
\Planck+BK18+ACT & $r$+$\beta^2_{F}$ & $< 3.35$ & -  & - & $< 4.91 $ \\
\Planck+BK18+ACT & $r$+$\beta^2_{AF,T}$+$\beta^2_{AF,S}$+$\beta^2_{F}$ & $< 3.03$ &  $< 0.0655$ & $< 0.0645$ & $< 4.76 $ \\
\hline
\end{tabular}
\caption{Bounds at 95\% CL on $r$, $\beta^2_{AF,T}$, $\beta^2_{AF,S}$, $\beta^2_{F}$ for the listed datasets and models. Eqs. \eqref{eq:clTE}-\eqref{eq:K33} show how the $\beta^2$ parameters affect the CMB spectra.
The limits have been expressed in units of $10^{-2}$. The key ``VV'' represents the combined CLASS+SPIDER dataset for V-modes.}\label{tab:beta_results}
\end{table}

\section{Implications for the LV coefficients in the minimal SME action} \label{sec:Connection}
In this Section, we translate the bounds on the phenomenological parameters $\beta^2_{AF,T}$, $\beta^2_{AF,S}$ and $\beta^2_{F}$ introduced in Eqs.~\eqref{eq:beta_0},~\eqref{eq:beta_123} and \eqref{eq:beta-F} into constraints on the LV couplings $k_{AF}$ and $k_F$ appearing in the action in Eq.~\eqref{eq:full-L-FLRW}.
We focus on the constraints obtained with the full dataset combination, \Planck+BK18+ACT. 
We report these results in Tab.~\ref{tab:LV_coeff_results}.

Focussing first on the CPT-odd effects, the constraints on the time component of $k_{AF}$ are usually rephrased as bounds on the parameter $k_{(V)00}^{(3)} = -\sqrt{4\pi} (k_{AF})^0$ (see Refs.~\cite{Kostelecky:2008ts,Kostelecky:2008be}).
This parameter can be linked to the phenomenological parameter $\beta^2_{AF,T}$ as follows:
\begin{equation}\label{eq:k00}
    |k_{(V)00}^{(3)}| = \frac{\sqrt{\pi}}{2c(\tau_0 - \tau_{\rm LS})} \sqrt{\beta^2_{AF,T}} \simeq 4.09 \times 10^{-43} \sqrt{\beta^2_{AF,T}} \; {\rm GeV} \, ,
\end{equation}
where we have assumed that $(k_{AF})_0$ is constant along the line of sight and
\begin{equation}\label{eq:ctau}
    c(\tau_0 - \tau_{\rm LS}) = \frac{c}{H_0}\int_0^{z_{\rm LS}} \frac{dz}{\left[\Omega_r(1+z)^4 + \Omega_m(1+z)^3 + \Omega_\Lambda\right]^{1/2}} \simeq 13870 \, {\rm Mpc} \, .
\end{equation}
In order to get the estimate in Eq.~\eqref{eq:ctau}, we have used the best-fit values for the cosmological parameters taken from \Planck\ 2018 (TT, TE, EE + lowE constraints for $\Lambda$CDM model)~\cite{planck2016-l06}.

Analogously, from Eq.~\eqref{eq:beta_123} we find for the space components of $k_{AF}$
\begin{equation}\label{eq:kAF}
    |\mathbf{k_{AF}}| \simeq 2.00 \times 10^{-43} \sqrt{\beta^2_{AF,S}} \; {\rm GeV} \, .
\end{equation}

For what concerns the CPT-even effects, recasting our constraints on $\beta^2_F$ into bounds on the components of $k_F$ is less trivial, due to the frequency dependence of Eqs.~\eqref{eq:beta_FE}-\eqref{eq:beta_FB}. 
From Eqs.~\eqref{eq:beta_FE}-\eqref{eq:beta_FB} we obtain 
\begin{equation}
    \label{eq:kF}
    k_{F,E+B} \equiv \left(2k_{F,E}^2 + \frac{k_{F,B}^2}{3}\right)^{1/2} \simeq 4.38 \times 10^{-29} \left(\frac{\nu}{{\rm GHz}}\right)^{-1} \sqrt{\beta^2_F} \, .
\end{equation}
To account for the fact that we are combining information coming from different experiments, observing the sky in different frequency channels, we can define an effective frequency $\nu_f$ following the method presented in Ref.~\cite{Gubitosi:2012rg}. Given the frequency dependence in Eq.~\eqref{eq:kF}, we find 
\begin{equation}
    \label{eq:effective-frequency}
    \nu_f = \left(\frac{\sum_i\frac{1}{\sigma_i^2}\left[\ln\left(\frac{\nu_+^i}{{\rm GHz}}\right)-\ln\left(\frac{\nu_-^i}{{\rm GHz}}\right)\right
    ]}{\sum_i\frac{1}{\sigma_i^2}\left(\frac{\nu_+^i}{{\rm GHz}}-\frac{\nu_-^i}{{\rm GHz}}\right)}\right)^{-1} \, {\rm GHz} \, ,
\end{equation}
where $\left[\nu_-^i,\nu_+^i\right]$ is the frequency interval of the $i$-th frequency channel and $\sigma_i$ is the noise level. 
Using Eq.~\eqref{eq:effective-frequency}, we obtain $\nu_f = 158.8 \, {\rm GHz}, 121.7 \, {\rm GHz}$ and $122.7 \, {\rm GHz}$ for Planck~\cite{planck2016-l01,planck2016-l03}, BK18~\cite{BICEPKeck:2020lwr,BICEP:2021xfz} and ACT~\cite{Thornton:2016wjq,ACT:2020gnv}, respectively. 

We now report the 95\% CL constraints on the LV coefficients using \Planck+BK18+ACT data, in the case where the three parameters $\beta^2_{AF,T}$, $\beta^2_{AF,S}$ and $\beta^2_F$ are all free to vary. For the CPT-odd terms we find
\begin{align}
    \label{eq:bounds-CPTodd-time}
    |k_{(V)00}^{(3)}| &< 1.05 \times 10^{-44} \; {\rm GeV} \, , \\
    \label{eq:bounds-CPTodd-space}
    |\mathbf{k_{AF}}| &< 0.51 \times 10^{-44} \; {\rm GeV} \, ,
\end{align}
whereas for the CPT-even operator we obtain
\begin{equation}
    \label{eq:bound-E+B}
    k_{F,E+B} < 6.02 \times 10^{-32} \left(\frac{\nu_f}{158.8 \, {\rm GHz}}\right)^{-1} \, .
\end{equation}
Note that the bound on $k_{F,E+B}$ in Eq.~\eqref{eq:bound-E+B} has been obtained by normalizing to the effective frequency of \Planck, $\nu_f = 158.8$ GHz, since the constraint on $\beta^2_F$ is driven by the rescaling of the TE and EE spectra, dominated by \Planck\ data (see Sec.~\ref{sec:Results}).
We remind the reader that the full set of constraints derived from different data and parameter combinations can be found in Table \ref{tab:LV_coeff_results}.

The bounds on the LV coefficients derived in previous literature are collected in \cite{Kostelecky:2008ts}, see Tables D15 and D16.
For the CPT-odd case, an upper bound on the parameter $|k_{(V)00}^{(3)}|$ has been obtained in Ref.~\cite{Kahniashvili:2008va} using WMAP data, leading to the result $|k_{(V)00}^{(3)}|< 4.9 \times 10^{-43} \; {\rm GeV}$ at 95\% CL. We note that the limit derived in our analysis using \Planck+BK18+ACT data is stronger by more than one order of magnitude, see Eq.~\eqref{eq:bounds-CPTodd-time}. Analogously, a limit on the coefficient $|\mathbf{k_{\rm AF}}|$ from WMAP data has been obtained in \cite{Kostelecky:2007zz,Kostelecky:2008be}, yielding $|\mathbf{k_{\rm AF}}| < 2 \times 10^{-42} \; {\rm GeV}$ at 95\% CL. In this case, the bound derived in our analysis is stronger by two orders of magnitude, see Eq.~\eqref{eq:bounds-CPTodd-space}. We stress that the bounds on the CPT-odd coefficients derived in this work are the strongest to date, both considering CMB and other sources. See again Ref.~\cite{Kostelecky:2008ts} for an exhaustive list of current bounds.

For what concerns CPT-even Lorentz violation, our bound on $k_{F,E+B}$ improves previous constraints by more than one order of magnitude~\cite{Kostelecky:2007zz}.
The CMB-based cosmological bounds on the CPT-even coefficients are only overcome by those obtained from optical polarimetry of extragalactic sources, see Refs.~\cite{Friedman:2020bxa,Gerasimov:2021chj,Kostelecky:2008ts}.

The bounds presented in the previous paragraphs are obtained in the most general case with all the $\beta^2$ parameters jointly varied. This represents a further novelty of our work. However, it is worth mentioning that, since all the parameters compete for the same power, the bounds obtained with a single parameter exploration are slightly weaker, as can be seen in Table~\ref{tab:LV_coeff_results}. 

\begin{table}
\centering 
\begin{tabular}{l l c c c} 
\hline 
\rule{0pt}{11pt}
\multirow{2}{*}{Dataset} & \multirow{2}{*}{Model (\lcdm+)} &  $|k_{(V)00}^{(3)}|\times10^{44}$ &  $|\mathbf{k_{AF}}|\times10^{44}$ & \multirow{2}{*}{$k_{F,E+B}\times10^{32}$} \\
 & & (GeV) & (GeV) & \\
\hline
\hline 
\rule{0pt}{11pt}
\hspace{-1.2mm}\Planck & $\beta^2_{AF,T}$+$\beta^2_{AF,S}$ & $< 4.65$ & $< 2.26$  & - \\
\Planck & $r$+$\beta^2_{AF,T}$+$\beta^2_{AF,S}$  & $< 4.06$ & $< 1.95$  & - \\
\Planck+BK18 & $r$+$\beta^2_{AF,T}$ & $< 1.17$  & - & - \\
\Planck+BK18 & $r$+$\beta^2_{AF,S}$ & -  & $< 0.57$ & - \\
\Planck+BK18 & $r$+$\beta^2_{AF,T}$+$\beta^2_{AF,S}$ & $< 1.06$  & $< 0.53$ & - \\
\Planck+BK18 & $r$+$\beta^2_{F}$ & -  & - & $< 6.02$ \\
\Planck+BK18 & $r$+$\beta^2_{AF,T}$+$\beta^2_{AF,S}$+$\beta^2_{F}$ & $< 1.06$ & $< 0.53$ & $< 5.92$ \\
\Planck+BK18+ACT & $r$+$\beta^2_{AF,T}$ & $ < 1.13$ & - & - \\
\Planck+BK18+ACT & $r$+$\beta^2_{AF,S}$ & -  & $< 0.55$ & - \\
\Planck+BK18+ACT & $r$+$\beta^2_{AF,T}$+$\beta^2_{AF,S}$ & $< 1.05$ & $< 0.52$ & - \\
\Planck+BK18+ACT & $r$+$\beta^2_{F}$ & -  & - & $< 6.11$ \\
\Planck+BK18+ACT & $r$+$\beta^2_{AF,T}$+$\beta^2_{AF,S}$+$\beta^2_{F}$ &  $< 1.05$ & $< 0.51$ & $< 6.02$ \\
\hline
\end{tabular}
\caption{Bounds at 95\% CL on $k_{(V)00}^{(3)}$, $|\mathbf{k_{AF}}|$ and $k_{F,E+B}$ for the listed datasets and models. 
The constraints on $k_{F,E+B}$ are derived taking the effective frequency of \Planck, $\nu_f = 158.8$ GHz, since the constraint on $\beta^2_F$ is driven by \Planck\ TE and EE data.
}\label{tab:LV_coeff_results}
\end{table}

\section{Conclusions}
In this paper, we have derived the signatures of Lorentz-violating (LV) electrodynamics on the polarization of the cosmic microwave background (CMB) and provided the most stringent constraints to date on LV coefficients from CMB observations.
We computed the modified CMB spectra, employing the full expression of the LV action given in Eq.~\eqref{eq:full-L-FLRW}, and we performed a likelihood analysis exploiting the most recent CMB datasets.
To our knowledge, this is the first time that such an end-to-end analysis has been performed.
We considered the minimal Standard Model extension of electrodynamics, including both CPT-odd (mass dimension $d=3$) and CPT-even (mass dimension $d=4$) operators. The CPT-odd operator, characterized by the 4-vector $(k_{AF})_\mu$, is responsible for the standard cosmic birefringence effect (isotropic and anisotropic). The CPT-even operator, instead, is characterized by a tensor $(k_F)^{\mu \nu \rho \sigma}$ and converts linear into circular polarization, giving rise to a non-zero V-mode spectrum. 

The expressions for the modified CMB spectra are presented in Eqs.~\eqref{eq:clTE}-\eqref{eq:K33} and are obtained following the formalism laid down in Ref.~\cite{Lembo:2020ufn}. The LV effects are encoded in four phenomenological parameters, defined in Eqs.~\eqref{eq:beta_0}-\eqref{eq:beta_FB}.
The parameters characterizing the CPT-odd term are $\beta^2_{AF,T}$ and $\beta^2_{AF,S}$, related to the time and space components of $k_{AF}$, respectively. The CPT-even terms are $\beta^2_{F,E}$ and $\beta^2_{F,B}$, which depend on the components of the tensor $k_F$.
The theoretical predictions of the modified CMB spectra in presence of LV effects are computed by using a customized version of the Boltzmann solver \texttt{camb}.\footnote{The modified version of \texttt{camb} is available at \url{https://github.com/sgiardie/CAMB_CPT}.}
  
We derived constraints on the phenomenological LV parameters from state-of-the-art CMB datasets: \Planck~\citep{planck2016-l01}, BK18~\citep{BICEP:2021xfz}, ACT~\citep{ACT:2020gnv}, CLASS~\citep{Padilla:2019dhz} and SPIDER~\cite{SPIDER:2017kwu}. Table~\ref{tab:beta_results} shows the 95\% confidence intervals of the $\beta^2$ parameters, for different combinations of datasets and different choices of the underlying cosmological model. 
Sampling the LV coefficients does not affect significantly the standard cosmological parameters. 
The tensor-to-scalar ratio $r$ represents the only relevant exception.
Indeed, the constraint on $r$ is $\sim 10\%$ tighter when all $\beta^2$ parameters are sampled with respect to the \lcdm+$r$ model (see the discussion in Section \ref{sec:Results} and the full triangle plots in Appendix \ref{app:plots}). 
As far as the CPT-even term is concerned, we found that current V-mode datasets have negligible constraining power compared to measurements of linear CMB polarization, in agreement with previous findings (see Ref.~\cite{Lembo:2020ufn}). 

Finally, we recast the constraints on the phenomenological parameters $\beta^2$ into bounds on the coefficients of the CPT-even and -odd operators appearing in the minimal SME action, see Table \ref{tab:LV_coeff_results}. We compared the constraints derived in this work with previous bounds from astrophysical and laboratory probes available in literature\footnote{Note that our analysis considers LV renormalizable operators of dimensions 3 and 4, whose effects can be better probed at low energy, e.g. using CMB radiation. On the other hand, higher-order LV operators lead to modifications of the photon dispersion relation that are more relevant at higher photon energies. Therefore, they are better constrained using high-energy radiation sources, such as gamma ray bursts~\cite{Kahniashvili:2006dt} and active galactic nuclei~\cite{Friedman:2020bxa,Gerasimov:2021chj}. See the review \cite{Addazi:2021xuf} for a more complete account of these tests.}. 

Our constraints on the CPT-odd parameters, i.e. $|k_{(V)00}^{(3)}| < 1.05 \times 10^{-44} \; {\rm GeV}$ and $|\mathbf{k_{AF}}| < 0.51 \times 10^{-44} \; {\rm GeV}$, are roughly one and two orders of magnitude tighter than previous CMB limits, respectively. Moreover, they are the strongest bounds obtained to date on the CPT-odd LV coefficients considering all other probes. Concerning the CPT-even case, the bounds are currently dominated by the constraint coming from optical polarimetry of extragalactic sources. Nevertheless, we improve previous CMB-based results by more than one order of magnitude, yielding $k_{F,E+B} < 6.02 \times 10^{-32}$.

Forthcoming CMB experiments, such as LiteBIRD~\citep{LiteBIRD:2022cnt}, Simons Observatory~\citep{SimonsObservatory:2018koc} and CMB-Stage 4~\citep{CMB-S4:2016ple}, will largely improve our sensitivity on such extensions of the standard electrodynamics, thanks to unprecedented sensitivity to linear CMB polarization as well as better sensitivity to V-mode polarization. A rough estimate of the expected improvements can be obtained by conservatively assuming that the constraints on the $\beta^2$ parameters will still be dominated by B-mode measurements. Future CMB experiments will increase their sensitivity to the tensor-to-scalar ratio $r$ by more than a factor of twenty compared to current bounds. The improvement on $r$ can be then translated to the same improvement on each $\beta^2$, since both parameters act as a rescaling factor for the BB spectrum (see Eq.~\eqref{eq:BB}). From Eqs.~\eqref{eq:k00},\eqref{eq:kAF},\eqref{eq:kF}, it is straightforward to eventually forecast a factor of 5 improvement on the physical coefficients in the LV action. Note that this is a conservative estimate since it does not account for the increased constraining power coming from more accurate measurements of E-mode polarization. Improved V-mode bounds would also allow to disentangle the effects of the phenomenological $\beta^2_{F,E}$ and $\beta^2_{F,B}$ parameters. This would potentially set individual bounds on these two coefficients, whose effects are indistinguishable when exploiting measurements of linear polarization only, see discussion in Sec.~\ref{sec:Results}. A detailed forecast analysis is left as the subject of a future publication.

\acknowledgments

We thank Alessandro Gruppuso and Paolo Natoli for useful discussions while this paper was in preparation and feedback on the final version of the manuscript. We acknowledge financial support from the INFN InDark initiative and from the COSMOS network (www.cosmosnet.it) through the ASI (Italian Space Agency) Grants 2016-24-H.0 and 2016-24-H.1-2018, as well as 2020-9-HH.0 (participation in LiteBIRD phase A). SG acknowledges postdoctoral support from the European Research Council (ERC) under the European Union’s Horizon 2020 research and innovation programme (Grant agreement No. 849169). GG acknowledges Perimeter Institute for hospitality in December 2022, when this project was completed. Research at Perimeter Institute for Theoretical Physics is supported in part by the Government of Canada through NSERC and by the Province of Ontario through MRI. GG also acknowledges participation in the COST Action CA18108 “Quantum gravity phenomenology in the multi-messenger approach”.
We acknowledge the use of \texttt{numpy} \citep{Harris:2020xlr}, \texttt{matplotlib} \citep{Hunter:2007ouj} and \texttt{getdist} \citep{Lewis:2019xzd} software packages, and the use of computing facilities at CINECA.

\bibliographystyle{JHEP}
\bibliography{bibliography,Planck_bib}

\appendix
\section{Plot appendix} \label{app:plots}
For completeness, we collect here the full triangle plots for all the cases discussed in Sec.~\ref{sec:Results} of the main text. The triangle plots reported in this appendix include all the cosmological parameters sampled in the MCMC analysis, as detailed in Sec.~\ref{sec:method}. Apart from the correlations already discussed at length in the main text (see Sec.~\ref{sec:Results}), the inclusion of the $\beta$ parameters in the analysis does not lead to significant modifications of the posterior distributions of the remaining cosmological parameters with respect to the standard (i.e., no LV) scenario. The shifts in some of the posterior distributions observed when including ACT data in the analysis are known features not specific to this work and have been discussed at length in the relevant ACT publications, see e.g.,~\cite{ACT:2020gnv,ACT:2020frw}  

\begin{figure}
\centering
\includegraphics[width=0.8\textwidth]{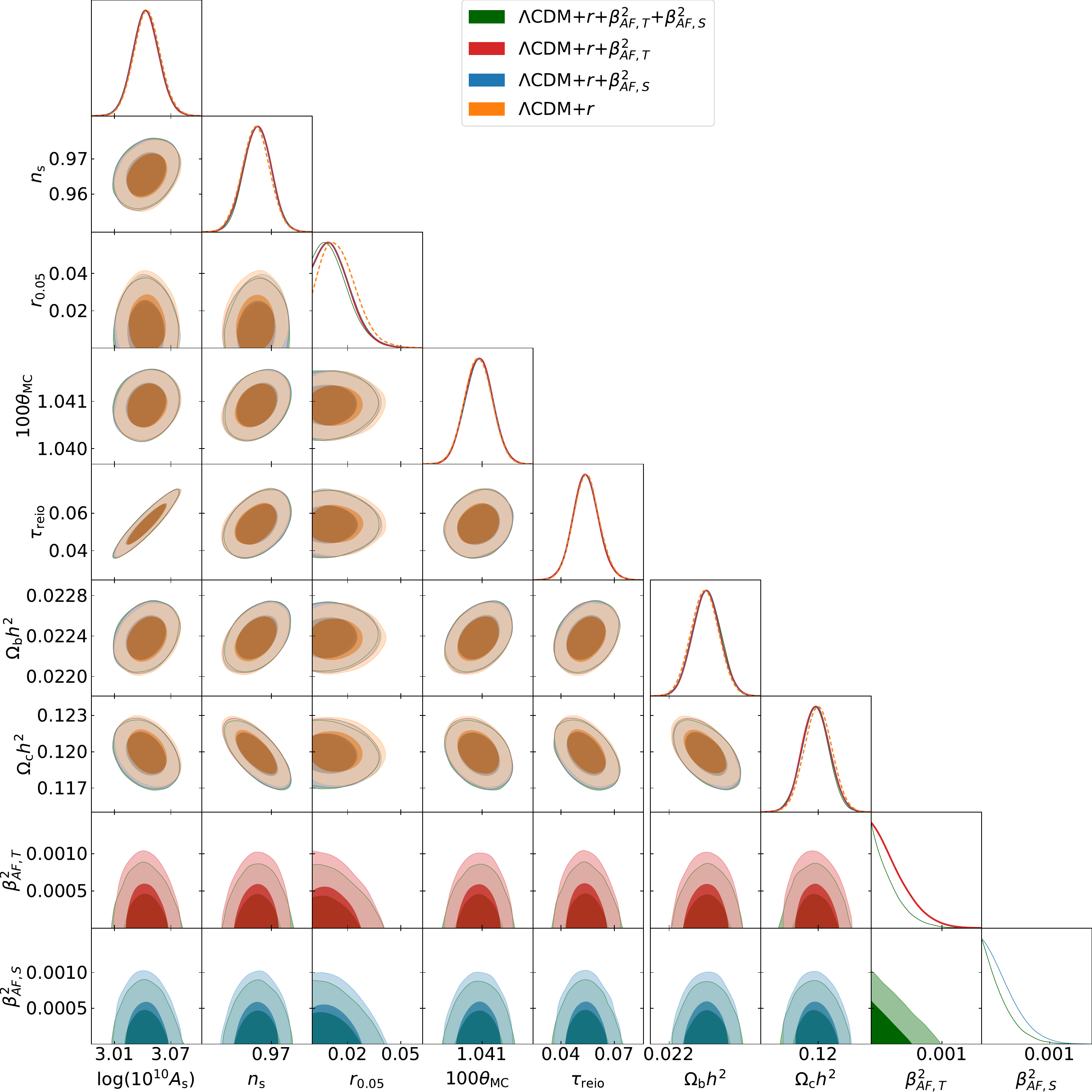}
\caption{One and two-dimensional posterior probability distribution for the full set of parameters varied in the MCMC analysis. We report the constraints obtained when assuming $\Lambda$CDM+$r$+$\beta^2_{AF,T}$+$\beta^2_{AF,S}$ (in green), $\Lambda$CDM+$r$+$\beta^2_{AF,T}$ (in red), $\Lambda$CDM+$r$+$\beta^2_{AF,S}$ (in blue)  and $\Lambda$CDM+$r$ (in orange) using the \Planck\ TTTEEE+lensing+BK18 dataset. 
The marginalization over either $\beta^2_{AF,T}$ or $\beta^2_{AF,S}$ has the same effect on the other parameters. It is worth to underline the tighter limit on $r$ with respect to the case in which each $\beta^2_{AF}$ is equal to zero. Opening to both the $\beta^2_{AF}$ parameters shrinks even more the constraints on $\beta^2_{AF,T}$, $\beta^2_{AF,S}$ and $r$.} \label{fig:lcdm+r+beta_AF_complete}
\end{figure}

\begin{figure}
\centering
\includegraphics[width=0.8\textwidth]{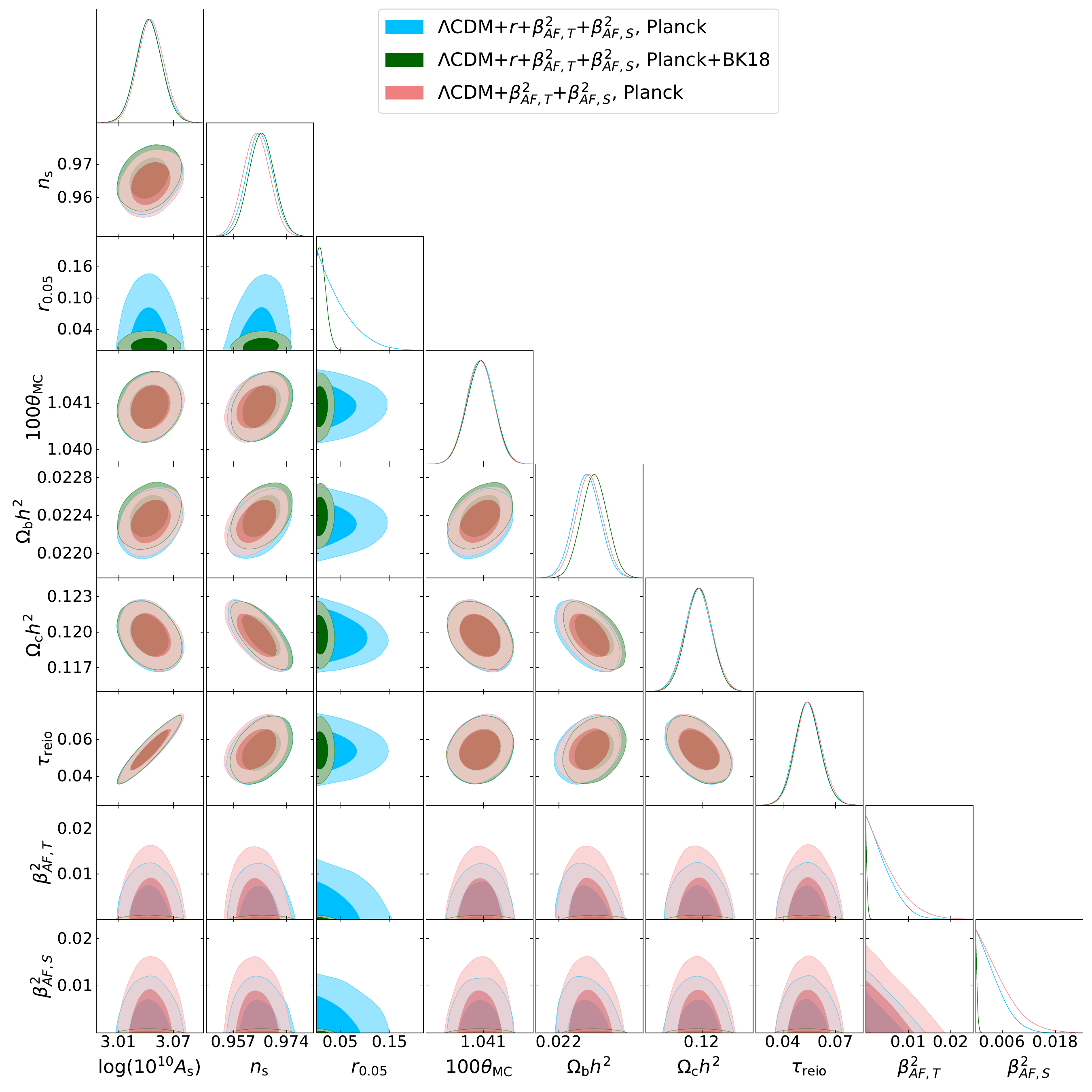}
\caption{One and two-dimensional posterior probability distribution for the full set of parameters varied in the MCMC analysis. We report the constraints obtained when assuming $\Lambda$CDM+$\beta^2_{AF,T}+\beta^2_{AF,S}$ (in pink) and $\Lambda$CDM+$r$+$\beta^2_{AF,T}+\beta^2_{AF,S}$ (in cyan and green). The former using only \Planck\ TTTEEE+lensing dataset, while the latter using both \Planck\ TTTEEE+lensing and \Planck\ TTTEEE+lensing+BK18 datasets. 
Note how much the constraints on the $\beta^2_{AF}$ parameters shrink when marginalizing over $r$ and adding the BK18 dataset. The shift on $n_s$ and $\Omega_b h^2$ is instead due especially to the change of dataset.} \label{fig:lcdm+beta_AF_Plonly_vs_PlBK18_complete}
\end{figure}

\begin{figure}
\centering
\includegraphics[width=0.8\textwidth]{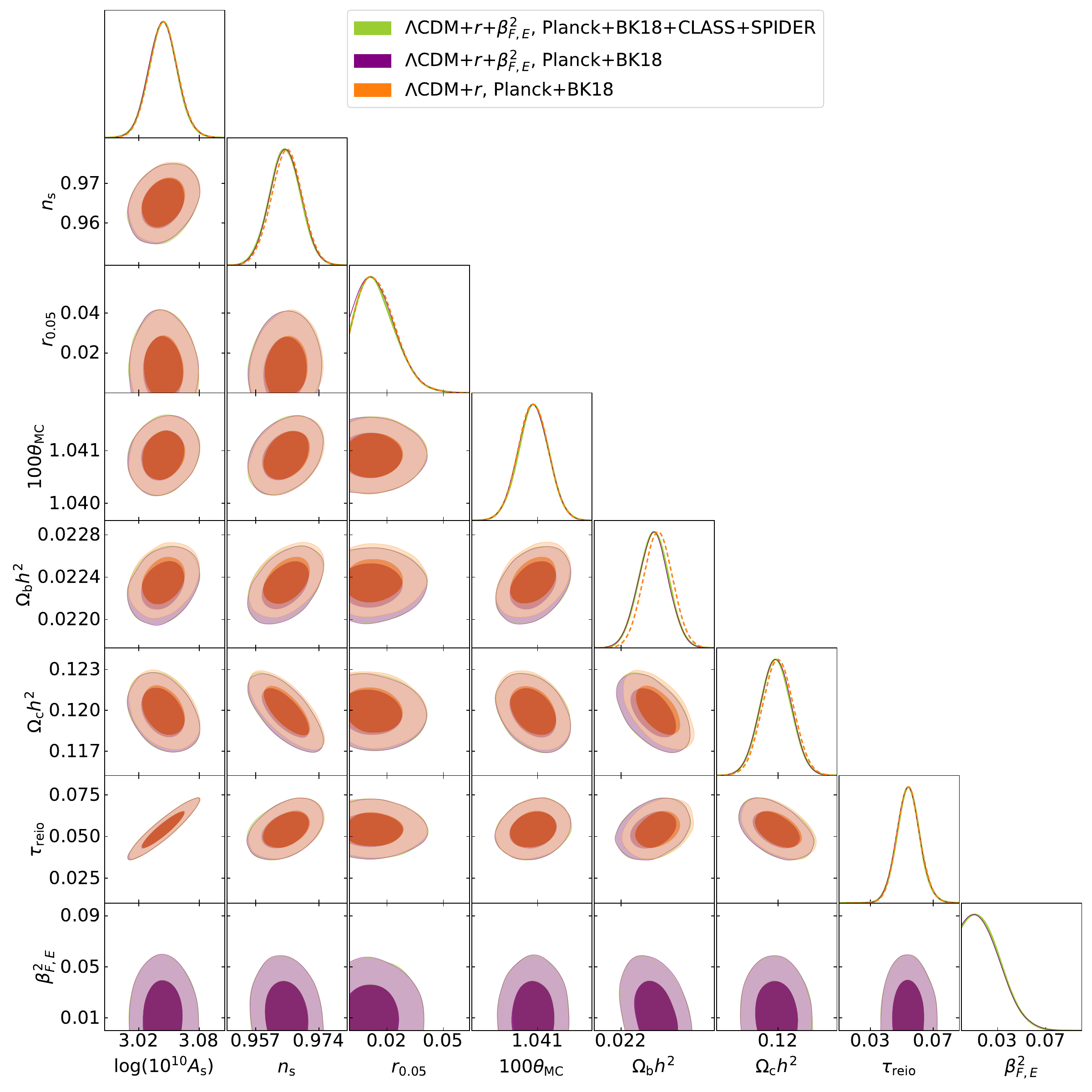}
\caption{One and two-dimensional posterior probability distribution for the full set of parameters varied in the MCMC analysis. We report the constraints obtained when assuming the $\Lambda$CDM+$r$+$\beta^2_{F,E}$ model using the Planck TTTEEE+lensing+BK18+CLASS+SPIDER dataset (in lime) and assuming $\Lambda$CDM+$r$+$\beta^2_{F,E}$(in purple) and $\Lambda$CDM+$r$ (in orange), both using the Planck TTTEEE+lensing+BK18 dataset.
The cases $\Lambda$CDM+$r$+$\beta^2_{F,E}$ with and without V-modes data are perfectly overlapping, showing the lack of constraining power from the current circular polarization data and justifying the choice of sampling over the combination $\beta^2_F$ in Eq. \eqref{eq:beta-F}} \label{fig:lcdm+r+beta_F_complete}
\end{figure}

\begin{figure}
\centering
\includegraphics[width=0.8\textwidth]{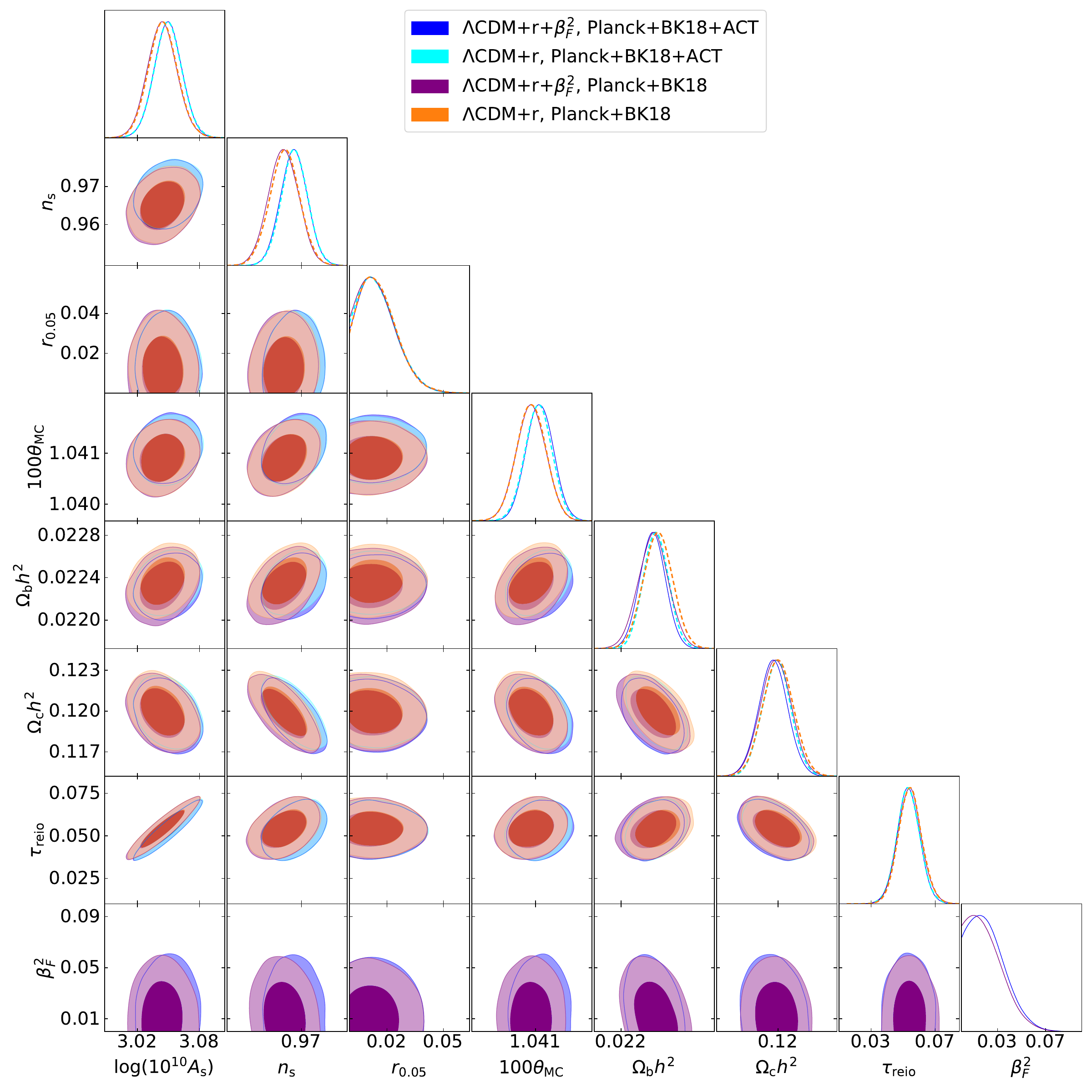}
\caption{One and two-dimensional posterior probability distribution for the full set of parameters varied in the MCMC analysis. We report the constraints obtained when assuming the $\Lambda$CDM+$r$+$\beta^2_{F}$ model using the Planck TTTEEE+lensing+BK18 dataset (in orange) and Planck TTTEEE+lensing+BK18+ACT (in blue) and when assuming the $\Lambda$CDM+$r$ model using Planck TTTEEE+lensing+BK18 (in dashed orange) and Planck TTTEEE+lensing+BK18+ACT (in dashed cyan). Notice the shifts in $\Omega_{b/c} h^2$ due to the sampling of $\beta^2_F$. The addition of ACT data widens the limit on $\beta^2_F$, due to the preference of ACT data for higher $A_s$ and $n_s$.} \label{fig:lcdm+r+beta_F_act_complete}
\end{figure}

\begin{figure}
\centering
\includegraphics[width=0.8\textwidth]{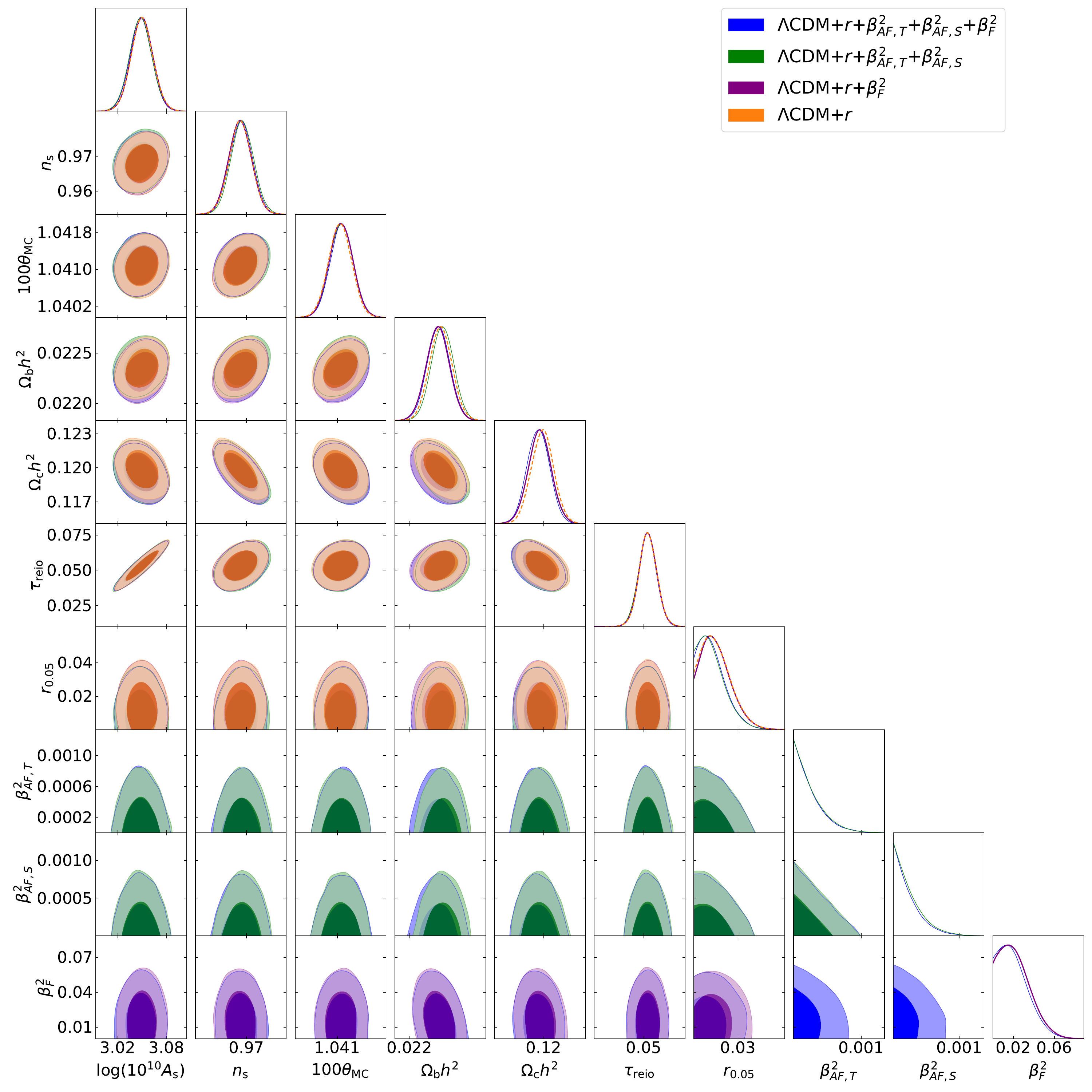}
\caption{One and two-dimensional posterior probability distribution for the full set of parameters varied in the MCMC analysis. We report the constraints obtained when assuming $\Lambda$CDM+$r$+$\beta^2_{AF,T}$+$\beta^2_{AF,S}$+$\beta^2_{F}$ (in dark blue), $\Lambda$CDM+$r$+$\beta^2_{AF,T}$+$\beta^2_{AF,S}$ (in green), $\Lambda$CDM+$r$+$\beta^2_{F}$ (in purple) and $\Lambda$CDM+$r$ (in orange), using the \Planck\ TTTEEE+lensing+BK18+ACT dataset.
Varying all the $\beta^2$ parameters together shrinks the constraints on $\beta^2_F$, leaving almost unchanged those on $r$ and the $\beta^2_{AF}$ parameters. Note the same shifts on $\Omega_b h^2$, $\Omega_c h^2$ in both the blue and the purple case, driven by the marginalization over $\beta^2_F$.} \label{fig:lcdm+r+beta_complete}
\end{figure}

\begin{figure}
\centering
\includegraphics[width=0.7\textwidth]{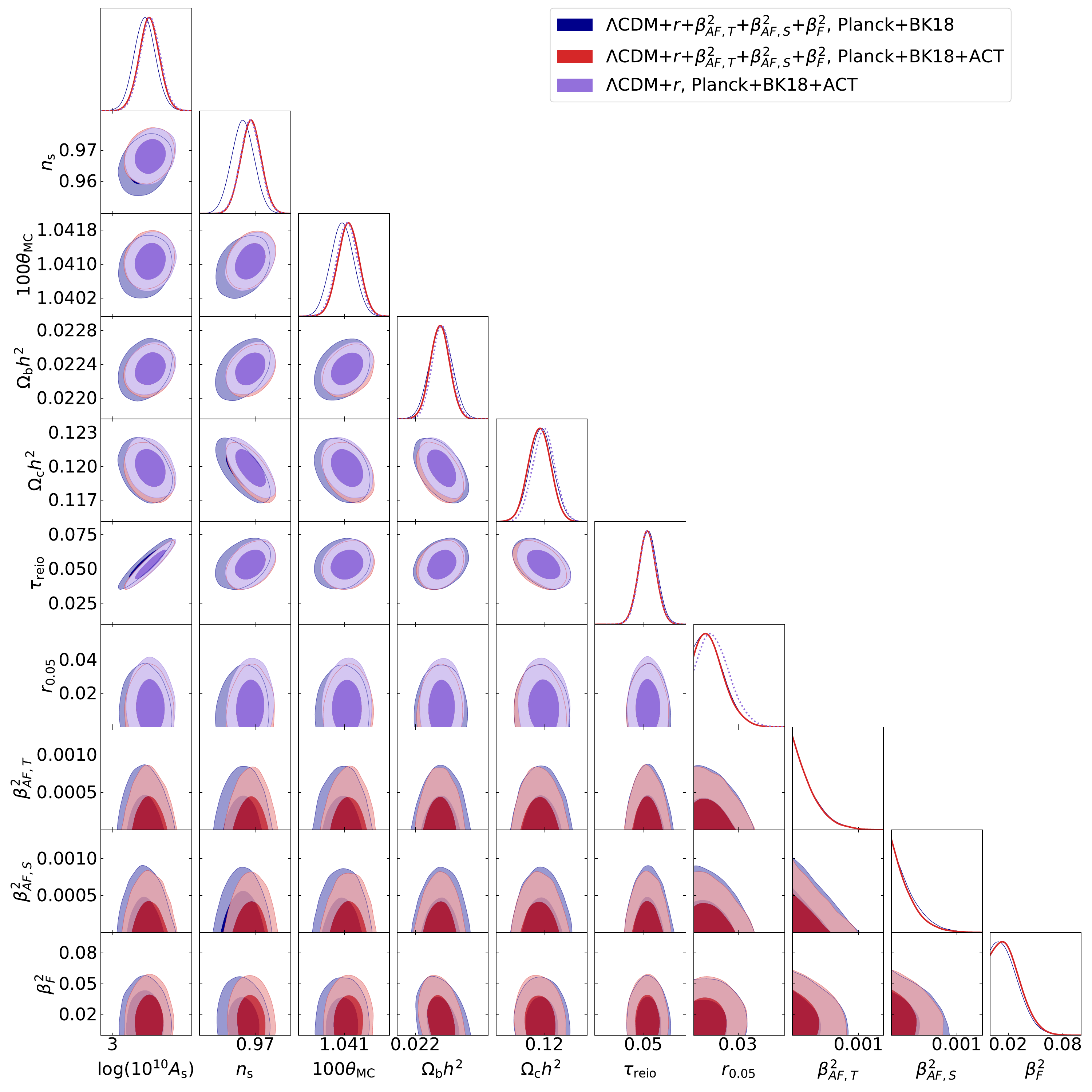}
\caption{One and two-dimensional posterior probability distribution for the full set of parameters varied in the MCMC analysis. We report the constraints obtained when assuming the $\Lambda$CDM+$r$+$\beta^2_{AF,T}$+$\beta^2_{AF,S}$+$\beta^2_{F}$ model using \Planck\ TTTEEE+lensing+BK18 (in dark blue) and \Planck\ TTTEEE+lensing+BK18+ACT (in red) datasets. For comparison, we also show the constrain obtained assuming a $\Lambda$CDM+$r$ model using the \Planck\ TTTEEE+lensing+BK18+ACT dataset (in violet).
Including ACT data in the analysis degrades the constraints on $\beta^2_F$. The shifts between the blue and red cases in some cosmological parameters (such as $n_s$, $\theta_{\rm{MC}}$ and $A_s$) are due to the addition of ACT data. They are indeed also present in the violet case. This shift can be also noticed in Figure 17 of Ref.~\citep{ACT:2020gnv}.} \label{fig:lcdm+r+beta}
\end{figure}

\end{document}

%% file: Planck.tex
\def\setsymbol#1#2{\expandafter\def\csname #1\endcsname{#2}}
\def\getsymbol#1{\csname #1\endcsname}

\def\Planck{\textit{Planck}}

\newbox\tablebox    \newdimen\tablewidth
\def\leaderfil{\leaders\hbox to 5pt{\hss.\hss}\hfil}
\def\tablenote#1 #2\par{\begingroup \parindent=0.8em
    \abovedisplayshortskip=0pt\belowdisplayshortskip=0pt
    \noindent
    $$\hss\vbox{\hsize\tablewidth \hangindent=\parindent \hangafter=1 \noindent
    \hbox to \parindent{$^#1$\hss}\strut#2\strut\par}\hss$$
    \endgroup}

\def\L2{\ifmmode L_2\else $L_2$\fi}

\def\DeltaT{\ifmmode \Delta T\else $\Delta T$\fi}
\def\deltat{\ifmmode \Delta t\else $\Delta t$\fi}
\def\fknee{\ifmmode f_{\rm knee}\else $f_{\rm knee}$\fi}
\def\Fmax{\ifmmode F_{\rm max}\else $F_{\rm max}$\fi}
\def\solar{\ifmmode{\rm M}_{\mathord\odot}\else${\rm M}_{\mathord\odot}$\fi}
\def\Msolar{\ifmmode{\rm M}_{\mathord\odot}\else${\rm M}_{\mathord\odot}$\fi}
\def\Lsolar{\ifmmode{\rm L}_{\mathord\odot}\else${\rm L}_{\mathord\odot}$\fi}
\def\inv{\ifmmode^{-1}\else$^{-1}$\fi}
\def\mo{\ifmmode^{-1}\else$^{-1}$\fi}
\def\sup#1{\ifmmode ^{\rm #1}\else $^{\rm #1}$\fi}
\def\expo#1{\ifmmode \times 10^{#1}\else $\times 10^{#1}$\fi}
\def\,{\thinspace}
\def\lsim{\mathrel{\raise .4ex\hbox{\rlap{$<$}\lower 1.2ex\hbox{$\sim$}}}}
\def\gsim{\mathrel{\raise .4ex\hbox{\rlap{$>$}\lower 1.2ex\hbox{$\sim$}}}}

\def\simprop{\mathrel{\raise .4ex\hbox{\rlap{$\propto$}\lower 1.2ex\hbox{$\sim$}}}}
\def\deg{\ifmmode^\circ\else$^\circ$\fi}
\def\pdeg{\ifmmode $\setbox0=\hbox{$^{\circ}$}\rlap{\hskip.11\wd0 .}$^{\circ}
          \else \setbox0=\hbox{$^{\circ}$}\rlap{\hskip.11\wd0 .}$^{\circ}$\fi}
\def\arcs{\ifmmode {^{\scriptstyle\prime\prime}}
          \else $^{\scriptstyle\prime\prime}$\fi}
\def\arcm{\ifmmode {^{\scriptstyle\prime}}
          \else $^{\scriptstyle\prime}$\fi}
\newdimen\sa  \newdimen\sb
\def\parcs{\sa=.07em \sb=.03em
     \ifmmode \hbox{\rlap{.}}^{\scriptstyle\prime\kern -\sb\prime}\hbox{\kern -\sa}
     \else \rlap{.}$^{\scriptstyle\prime\kern -\sb\prime}$\kern -\sa\fi}
\def\parcm{\sa=.08em \sb=.03em
     \ifmmode \hbox{\rlap{.}\kern\sa}^{\scriptstyle\prime}\hbox{\kern-\sb}
     \else \rlap{.}\kern\sa$^{\scriptstyle\prime}$\kern-\sb\fi}
\def\ra[#1 #2 #3.#4]{#1\sup{h}#2\sup{m}#3\sup{s}\llap.#4}
\def\dec[#1 #2 #3.#4]{#1\deg#2\arcm#3\arcs\llap.#4}
\def\deco[#1 #2 #3]{#1\deg#2\arcm#3\arcs}
\def\rra[#1 #2]{#1\sup{h}#2\sup{m}}

\def\dots{\relax\ifmmode \ldots\else $\ldots$\fi}
\def\WHzsr{\ifmmode $W\,Hz\mo\,sr\mo$\else W\,Hz\mo\,sr\mo\fi}
\def\mHz{\ifmmode $\,mHz$\else \,mHz\fi}
\def\GHz{\ifmmode $\,GHz$\else \,GHz\fi}
\def\mKs{\ifmmode $\,mK\,s$^{1/2}\else \,mK\,s$^{1/2}$\fi}
\def\muKs{\ifmmode \,\mu$K\,s$^{1/2}\else \,$\mu$K\,s$^{1/2}$\fi}
\def\muKRJs{\ifmmode \,\mu$K$_{\rm RJ}$\,s$^{1/2}\else \,$\mu$K$_{\rm RJ}$\,s$^{1/2}$\fi}
\def\muKHz{\ifmmode \,\mu$K\,Hz$^{-1/2}\else \,$\mu$K\,Hz$^{-1/2}$\fi}
\def\MJysr{\ifmmode \,$MJy\,sr\mo$\else \,MJy\,sr\mo\fi}
\def\MJysrmK{\ifmmode \,$MJy\,sr\mo$\,mK$_{\rm CMB}\mo\else \,MJy\,sr\mo\,mK$_{\rm CMB}\mo$\fi}
\def\microns{\ifmmode \,\mu$m$\else \,$\mu$m\fi}

\def\muK{\ifmmode \,\mu$K$\else \,$\mu$\hbox{K}\fi}
\def\microK{\ifmmode \,\mu$K$\else \,$\mu$\hbox{K}\fi}
\def\muW{\ifmmode \,\mu$W$\else \,$\mu$\hbox{W}\fi}
\def\kms{\ifmmode $\,km\,s$^{-1}\else \,km\,s$^{-1}$\fi}
\def\kmsMpc{\ifmmode $\,\kms\,Mpc\mo$\else \,\kms\,Mpc\mo\fi}
\providecommand{\sorthelp}[1]{}

%% file: shortcuts.tex
\usepackage[dvipsnames]{xcolor}

\def\NHUNIT{\ifmmode {\rm \,cm^{-2}} \else $\rm \,cm^{-2}$ \fi} 

\def\muKcmb{\ifmmode \,\mu$K$_{\rm CMB}$\else \,$\mu$K$_{\rm CMB}$\fi}
\newcommand{\lcdm}{$\Lambda$CDM}

\newcommand{\OmegaM}{\ifmmode\Omega_{\rm M}\else $\Omega_{\rm M}$\fi}

\providecommand{\Planck}{\textit{Planck}}

\providecommand{\text}[1]{\rm{#1}}

\providecommand{\muK}{\mu\rm{K}}

\providecommand{\Cobaya}{{\tt Cobaya}}

\newcommand{\begm}{\begin{pmatrix}}
\newcommand{\enm}{\end{pmatrix}}

\def\pmb#1{\setbox0=\hbox{#1}%
    \kern-.025em\copy0\kern-\wd0
    \kern.05em\copy0\kern-\wd0
    \kern-.025em\raise.0433em\box0}

\def\p2Y{\;_2Y}
\def\m2Y{\;_{-2}Y}

\newcommand{\mksym}[1]{\ifmmode {\rm #1}\else #1\fi}

\providecommand{\text}[1]{\rm{#1}}

\providecommand{\muK}{\mu\rm{K}}

\newcommand\ba{\begin{eqnarray}}
\newcommand\ea{\end{eqnarray}}
\newcommand\bea{\begin{eqnarray}}
\newcommand\eea{\end{eqnarray}}

\newcommand\be{\begin{equation}}
\newcommand\ee{\end{equation}}



%% file: Planck_bib.bib
@ARTICLE{planck2016-l01,
author = {{\sorthelp{Planck Collaboration 2018A}}{Planck Collaboration I}},
title = "{\textit{Planck} 2018 results. I. Overview, and the cosmological
legacy of \textit{Planck}}",
journal = {\aap},
archivePrefix = "arXiv",
eprint = {1807.06205},
year = 2020,
volume = 641,
pages = {A1},
doi = {10.1051/0004-6361/201833880}
}

@ARTICLE{planck2016-l03, 
author = {{\sorthelp{Planck Collaboration 2018C}}{Planck Collaboration III}},
title = "{\textit{Planck} 2018 results. III. High Frequency Instrument data
 processing}",
journal = {\aap},
archivePrefix = "arXiv",
eprint = {1807.06207},
year = 2020,
volume = 641,
pages = {A3},
doi = {10.1051/0004-6361/201832909}
}

@ARTICLE{planck2016-l05, 
author = {{\sorthelp{Planck Collaboration 2018E}}{Planck Collaboration V}},
title = "{\textit{Planck} 2018 results. V. Power spectra and likelihoods}",
journal = {\aap},
archivePrefix = "arXiv",
eprint = {1907.12875},
year = 2020,
volume = 641,
pages = {A5},
doi = {10.1051/0004-6361/201936386}
}

@ARTICLE{planck2016-l06, 
author = {{\sorthelp{Planck Collaboration 2018F}}{Planck Collaboration VI}},
title = "{\textit{Planck} 2018 results. VI. Cosmological parameters}",
journal = {\aap},
archivePrefix = "arXiv",
eprint = {1807.06209},
year = 2020,
volume = 641,
pages = {A6},
doi = {10.1051/0004-6361/201833910}
}

@ARTICLE{planck2016-l08, 
author = {{\sorthelp{Planck Collaboration 2018H}}{Planck Collaboration VIII}},
title = "{\textit{Planck} 2018 results. VIII. Gravitational lensing}",
journal = {\aap},
archivePrefix = "arXiv",
eprint = {1807.06210},
year = 2020,
volume = 641,
pages = {A8},
doi = {10.1051/0004-6361/201833886}
}


%% file: bibliography.bib
@article{Campanelli:2004pm,
    author = "Campanelli, L. and Dolgov, A. D. and Giannotti, M. and Villante, F. L.",
    title = "{Faraday rotation of the CMB polarization and primordial magnetic field properties}",
    eprint = "astro-ph/0405420",
    archivePrefix = "arXiv",
    doi = "10.1086/424840",
    journal = "Astrophys. J.",
    volume = "616",
    pages = "1--7",
    year = "2004"
}

@article{Scoccola:2004ke,
    author = "Scoccola, Claudia and Harari, Diego and Mollerach, Silvia",
    title = "{B polarization of the CMB from Faraday rotation}",
    eprint = "astro-ph/0405396",
    archivePrefix = "arXiv",
    doi = "10.1103/PhysRevD.70.063003",
    journal = "Phys. Rev. D",
    volume = "70",
    pages = "063003",
    year = "2004"
}

@article{Pospelov:2008gg,
    author = "Pospelov, Maxim and Ritz, Adam and Skordis, Constantinos and Ritz, Adam and Skordis, Constantinos",
    title = "{Pseudoscalar perturbations and polarization of the cosmic microwave background}",
    eprint = "0808.0673",
    archivePrefix = "arXiv",
    primaryClass = "astro-ph",
    doi = "10.1103/PhysRevLett.103.051302",
    journal = "Phys. Rev. Lett.",
    volume = "103",
    pages = "051302",
    year = "2009"
}

@article{Carroll:1998zi,
    author = "Carroll, Sean M.",
    title = "{Quintessence and the rest of the world}",
    eprint = "astro-ph/9806099",
    archivePrefix = "arXiv",
    reportNumber = "NSF-ITP-98-063",
    doi = "10.1103/PhysRevLett.81.3067",
    journal = "Phys. Rev. Lett.",
    volume = "81",
    pages = "3067--3070",
    year = "1998"
}

@article{Carroll:1989vb,
    author = "Carroll, Sean M. and Field, George B. and Jackiw, Roman",
    title = "{Limits on a Lorentz and Parity Violating Modification of Electrodynamics}",
    reportNumber = "MIT-CTP-1782",
    doi = "10.1103/PhysRevD.41.1231",
    journal = "Phys. Rev. D",
    volume = "41",
    pages = "1231",
    year = "1990"
}

@article{Giovannini:2004pf,
    author = "Giovannini, Massimo",
    title = "{Magnetized birefringence and CMB polarization}",
    eprint = "hep-ph/0410387",
    archivePrefix = "arXiv",
    reportNumber = "CERN-PH-TH-2004-185",
    doi = "10.1103/PhysRevD.71.021301",
    journal = "Phys. Rev. D",
    volume = "71",
    pages = "021301",
    year = "2005"
}

@article{Balaji:2003sw,
    author = "Balaji, K. R. S. and Brandenberger, Robert H. and Easson, Damien A.",
    title = "{Spectral dependence of CMB polarization and parity}",
    eprint = "hep-ph/0310368",
    archivePrefix = "arXiv",
    reportNumber = "BROWN-HET-1380, MCGILL-22-03, SU-GP-03-10-2",
    doi = "10.1088/1475-7516/2003/12/008",
    journal = "JCAP",
    volume = "12",
    pages = "008",
    year = "2003"
}

@article{Liu:2006uh,
    author = "Liu, Guo-Chin and Lee, Seokcheon and Ng, Kin-Wang",
    title = "{Effect on cosmic microwave background polarization of coupling of quintessence to pseudoscalar formed from the electromagnetic field and its dual}",
    eprint = "astro-ph/0606248",
    archivePrefix = "arXiv",
    doi = "10.1103/PhysRevLett.97.161303",
    journal = "Phys. Rev. Lett.",
    volume = "97",
    pages = "161303",
    year = "2006"
}

@article{Finelli:2008jv,
    author = "Finelli, Fabio and Galaverni, Matteo",
    title = "{Rotation of Linear Polarization Plane and Circular Polarization from Cosmological Pseudo-Scalar Fields}",
    eprint = "0802.4210",
    archivePrefix = "arXiv",
    primaryClass = "astro-ph",
    doi = "10.1103/PhysRevD.79.063002",
    journal = "Phys. Rev. D",
    volume = "79",
    pages = "063002",
    year = "2009"
}

@article{Myers:2003fd,
    author = "Myers, Robert C. and Pospelov, Maxim",
    title = "{Ultraviolet modifications of dispersion relations in effective field theory}",
    eprint = "hep-ph/0301124",
    archivePrefix = "arXiv",
    reportNumber = "UVIC-TH-03-01, SUSX-TH-03-001",
    doi = "10.1103/PhysRevLett.90.211601",
    journal = "Phys. Rev. Lett.",
    volume = "90",
    pages = "211601",
    year = "2003"
}

@article{Galaverni:2014gca,
    author = "Galaverni, M. and Gubitosi, G. and Paci, F. and Finelli, F.",
    title = "{Cosmological birefringence constraints from CMB and astrophysical polarization data}",
    eprint = "1411.6287",
    archivePrefix = "arXiv",
    primaryClass = "astro-ph.CO",
    doi = "10.1088/1475-7516/2015/08/031",
    journal = "JCAP",
    volume = "08",
    pages = "031",
    year = "2015"
}

@article{Caldwell:2011pu,
    author = "Caldwell, Robert R. and Gluscevic, Vera and Kamionkowski, Marc",
    title = "{Cross-Correlation of Cosmological Birefringence with CMB Temperature}",
    eprint = "1104.1634",
    archivePrefix = "arXiv",
    primaryClass = "astro-ph.CO",
    doi = "10.1103/PhysRevD.84.043504",
    journal = "Phys. Rev. D",
    volume = "84",
    pages = "043504",
    year = "2011"
}

@article{Li:2008tma,
    author = "Li, Mingzhe and Zhang, Xinmin",
    title = "{Cosmological CPT violating effect on CMB polarization}",
    eprint = "0810.0403",
    archivePrefix = "arXiv",
    primaryClass = "astro-ph",
    doi = "10.1103/PhysRevD.78.103516",
    journal = "Phys. Rev. D",
    volume = "78",
    pages = "103516",
    year = "2008"
}

@article{Kamionkowski:2008fp,
    author = "Kamionkowski, Marc",
    title = "{How to De-Rotate the Cosmic Microwave Background Polarization}",
    eprint = "0810.1286",
    archivePrefix = "arXiv",
    primaryClass = "astro-ph",
    doi = "10.1103/PhysRevLett.102.111302",
    journal = "Phys. Rev. Lett.",
    volume = "102",
    pages = "111302",
    year = "2009"
}

@article{Addazi:2021xuf,
    author = "Addazi, A. and others",
    title = "{Quantum gravity phenomenology at the dawn of the multi-messenger era\textemdash{}A review}",
    eprint = "2111.05659",
    archivePrefix = "arXiv",
    primaryClass = "hep-ph",
    doi = "10.1016/j.ppnp.2022.103948",
    journal = "Prog. Part. Nucl. Phys.",
    volume = "125",
    pages = "103948",
    year = "2022"
}

@article{Amelino-Camelia:2008aez,
    author = "Amelino-Camelia, Giovanni",
    title = "{Quantum-Spacetime Phenomenology}",
    eprint = "0806.0339",
    archivePrefix = "arXiv",
    primaryClass = "gr-qc",
    doi = "10.12942/lrr-2013-5",
    journal = "Living Rev. Rel.",
    volume = "16",
    pages = "5",
    year = "2013"
}

@article{Mattingly:2005re,
    author = "Mattingly, David",
    title = "{Modern tests of Lorentz invariance}",
    eprint = "gr-qc/0502097",
    archivePrefix = "arXiv",
    doi = "10.12942/lrr-2005-5",
    journal = "Living Rev. Rel.",
    volume = "8",
    pages = "5",
    year = "2005"
}

@article{Gubitosi:2010dj,
    author = "Gubitosi, Giulia and Genovese, Giuseppe and Amelino-Camelia, Giovanni and Melchiorri, Alessandro",
    title = "{Planck-scale modifications to Electrodynamics characterizedh by a space-like symmetry-breaking vector}",
    eprint = "1003.0878",
    archivePrefix = "arXiv",
    primaryClass = "gr-qc",
    doi = "10.1103/PhysRevD.82.024013",
    journal = "Phys. Rev. D",
    volume = "82",
    pages = "024013",
    year = "2010"
}

@article{Torrado:2020dgo,
    author = "Torrado, Jesus and Lewis, Antony",
    title = "{Cobaya: Code for Bayesian Analysis of hierarchical physical models}",
    eprint = "2005.05290",
    archivePrefix = "arXiv",
    primaryClass = "astro-ph.IM",
    reportNumber = "TTK-20-15",
    doi = "10.1088/1475-7516/2021/05/057",
    journal = "JCAP",
    volume = "05",
    pages = "057",
    year = "2021"
}

@article{Lewis:1999bs,
      author         = "Lewis, Antony and Challinor, Anthony and Lasenby,
                        Anthony",
      title          = "{Efficient computation of CMB anisotropies in closed FRW
                        models}",
      journal        = "\apj",
      volume         = "538",
      year           = "2000",
      pages          = "473-476",
      doi            = "10.1086/309179",
      eprint         = "astro-ph/9911177",
      archivePrefix  = "arXiv",
      primaryClass   = "astro-ph",
      SLACcitation   = "%%CITATION = ASTRO-PH/9911177;%%"
}

@article{Howlett:2012mh,
    author = "Howlett, Cullan and Lewis, Antony and Hall, Alex and Challinor, Anthony",
    title = "{CMB power spectrum parameter degeneracies in the era of precision cosmology}",
    eprint = "1201.3654",
    archivePrefix = "arXiv",
    primaryClass = "astro-ph.CO",
    doi = "10.1088/1475-7516/2012/04/027",
    journal = "JCAP",
    volume = "04",
    pages = "027",
    year = "2012"
}

@article{Gubitosi:2014cua,
    author = "Gubitosi, Giulia and Martinelli, Matteo and Pagano, Luca",
    title = "{Including birefringence into time evolution of CMB: current and future constraints}",
    eprint = "1410.1799",
    archivePrefix = "arXiv",
    primaryClass = "astro-ph.CO",
    doi = "10.1088/1475-7516/2014/12/020",
    journal = "JCAP",
    volume = "12",
    pages = "020",
    year = "2014"
}

@article{Kostelecky:2007zz,
    author = "Kostelecky, V. Alan and Mewes, Matthew",
    title = "{Lorentz-violating electrodynamics and the cosmic microwave background}",
    eprint = "astro-ph/0702379",
    archivePrefix = "arXiv",
    reportNumber = "IUHET-504",
    doi = "10.1103/PhysRevLett.99.011601",
    journal = "Phys. Rev. Lett.",
    volume = "99",
    pages = "011601",
    year = "2007"
}

@article{Kostelecky:2009zp,
    author = "Kostelecky, V. Alan and Mewes, Matthew",
    title = "{Electrodynamics with Lorentz-violating operators of arbitrary dimension}",
    eprint = "0905.0031",
    archivePrefix = "arXiv",
    primaryClass = "hep-ph",
    reportNumber = "IUHET-527",
    doi = "10.1103/PhysRevD.80.015020",
    journal = "Phys. Rev. D",
    volume = "80",
    pages = "015020",
    year = "2009"
}

@article{Padilla:2019dhz,
doi = {10.3847/1538-4357/ab61f8},
url = {https://dx.doi.org/10.3847/1538-4357/ab61f8},
year = {2020},
month = {jan},
publisher = {The American Astronomical Society},
volume = {889},
number = {2},
pages = {105},
author = {Ivan L. Padilla and Joseph R. Eimer and Yunyang Li and Graeme E. Addison and Aamir Ali and John W. Appel and Charles L. Bennett and Ricardo Bustos and Michael K. Brewer and Manwei Chan and David T. Chuss and Joseph Cleary and Jullianna Couto and Sumit Dahal and Kevin Denis and Rolando Dünner and Thomas Essinger-Hileman and Pedro Fluxá and Dominik Gothe and Saianeesh K. Haridas and Kathleen Harrington and Jeffrey Iuliano and John Karakla and Tobias A. Marriage and Nathan J. Miller and Carolina Núñez and Lucas Parker and Matthew A. Petroff and Rodrigo Reeves and Karwan Rostem and Robert W. Stevens and Deniz Augusto Nunes Valle and Duncan J. Watts and Janet L. Weiland and Edward J. Wollack and Zhilei Xu},
title = {Two-year Cosmology Large Angular Scale Surveyor (CLASS) Observations: A Measurement of Circular Polarization at 40 GHz},
journal = "Astrophys. J.",
eprint = "1911.00391",
archivePrefix = "arXiv",
primaryClass = "astro-ph.CO"
}

@article{SPIDER:2017kwu,
    author = "Nagy, J. M. and others",
    collaboration = "SPIDER",
    title = "{A New Limit on CMB Circular Polarization from SPIDER}",
    eprint = "1704.00215",
    archivePrefix = "arXiv",
    primaryClass = "astro-ph.CO",
    reportNumber = "FERMILAB-PUB-17-135-AE",
    doi = "10.3847/1538-4357/aa7cfd",
    journal = "Astrophys. J.",
    volume = "844",
    number = "2",
    pages = "151",
    year = "2017"
}

@article{Colladay:1998fq,
    author = "Colladay, Don and Kostelecky, V. Alan",
    title = "{Lorentz violating extension of the standard model}",
    eprint = "hep-ph/9809521",
    archivePrefix = "arXiv",
    reportNumber = "IUHET-359",
    doi = "10.1103/PhysRevD.58.116002",
    journal = "Phys. Rev. D",
    volume = "58",
    pages = "116002",
    year = "1998"
}

@article{Kostelecky:2008ts,
    author = "Kostelecky, V. Alan and Russell, Neil",
    title = "{Data Tables for Lorentz and CPT Violation}",
    eprint = "0801.0287",
    archivePrefix = "arXiv",
    primaryClass = "hep-ph",
    month = "1",
    year = "2008"
}

@article{Kostelecky:2008be,
    author = "Kostelecky, V. Alan and Mewes, Matthew",
    title = "{Astrophysical Tests of Lorentz and CPT Violation with Photons}",
    eprint = "0809.2846",
    archivePrefix = "arXiv",
    primaryClass = "astro-ph",
    reportNumber = "IUHET-516",
    doi = "10.1086/595815",
    journal = "Astrophys. J. Lett.",
    volume = "689",
    pages = "L1--L4",
    year = "2008"
}

@article{Lembo:2020ufn,
    author = "Lembo, Margherita and Lattanzi, Massimiliano and Pagano, Luca and Gruppuso, Alessandro and Natoli, Paolo and Forastieri, Francesco",
    title = "{Cosmic Microwave Background Polarization as a Tool to Constrain the Optical Properties of the Universe}",
    eprint = "2007.08486",
    archivePrefix = "arXiv",
    primaryClass = "astro-ph.CO",
    doi = "10.1103/PhysRevLett.127.011301",
    journal = "Phys. Rev. Lett.",
    volume = "127",
    number = "1",
    pages = "011301",
    year = "2021"
}

@article{Kostelecky:2002hh,
    author = "Kostelecky, V. Alan and Mewes, Matthew",
    title = "{Signals for Lorentz violation in electrodynamics}",
    eprint = "hep-ph/0205211",
    archivePrefix = "arXiv",
    reportNumber = "IUHET-449",
    doi = "10.1103/PhysRevD.66.056005",
    journal = "Phys. Rev. D",
    volume = "66",
    pages = "056005",
    year = "2002"
}

@article{Gubitosi:2012rg,
    author = "Gubitosi, G. and Paci, F.",
    title = "{Constraints on cosmological birefringence energy dependence from CMB polarization data}",
    eprint = "1211.3321",
    archivePrefix = "arXiv",
    primaryClass = "astro-ph.CO",
    doi = "10.1088/1475-7516/2013/02/020",
    journal = "JCAP",
    volume = "02",
    pages = "020",
    year = "2013"
}

@article{BICEP:2021xfz,
    author = "Ade, P. A. R. and others",
    collaboration = "BICEP, Keck",
    title = "{Improved Constraints on Primordial Gravitational Waves using Planck, WMAP, and BICEP/Keck Observations through the 2018 Observing Season}",
    eprint = "2110.00483",
    archivePrefix = "arXiv",
    primaryClass = "astro-ph.CO",
    doi = "10.1103/PhysRevLett.127.151301",
    journal = "Phys. Rev. Lett.",
    volume = "127",
    number = "15",
    pages = "151301",
    year = "2021"
}

@article{Gelman:1992zz,
    author = "Gelman, Andrew and Rubin, Donald B.",
    title = "{Inference from Iterative Simulation Using Multiple Sequences}",
    doi = "10.1214/ss/1177011136",
    journal = "Statist. Sci.",
    volume = "7",
    pages = "457--472",
    year = "1992"
}

@article{ACT:2020gnv,
    author = "Aiola, Simone and others",
    collaboration = "ACT",
    title = "{The Atacama Cosmology Telescope: DR4 Maps and Cosmological Parameters}",
    eprint = "2007.07288",
    archivePrefix = "arXiv",
    primaryClass = "astro-ph.CO",
    doi = "10.1088/1475-7516/2020/12/047",
    journal = "JCAP",
    volume = "12",
    pages = "047",
    year = "2020"
}

@article{BICEPKeck:2020lwr,
    author = "St Germaine, T. and others",
    collaboration = "BICEP/Keck",
    title = "{Optical characterization of the Keck Array and BICEP3 CMB Polarimeters from 2016 to 2019}",
    eprint = "2002.05197",
    archivePrefix = "arXiv",
    primaryClass = "astro-ph.IM",
    doi = "10.1007/s10909-020-02392-8",
    journal = "J. Low Temp. Phys.",
    volume = "199",
    number = "3-4",
    pages = "824--832",
    year = "2020"
}

@article{Thornton:2016wjq,
    author = "Thornton, R. J. and others",
    title = "{The Atacama Cosmology Telescope: The polarization-sensitive ACTPol instrument}",
    eprint = "1605.06569",
    archivePrefix = "arXiv",
    primaryClass = "astro-ph.IM",
    doi = "10.3847/1538-4365/227/2/21",
    journal = "Astrophys. J. Suppl.",
    volume = "227",
    number = "2",
    pages = "21",
    year = "2016"
}

@article{Kahniashvili:2008va,
    author = "Kahniashvili, Tina and Durrer, Ruth and Maravin, Yurii",
    title = "{Testing Lorentz Invariance Violation with WMAP Five Year Data}",
    eprint = "0807.2593",
    archivePrefix = "arXiv",
    primaryClass = "astro-ph",
    doi = "10.1103/PhysRevD.78.123009",
    journal = "Phys. Rev. D",
    volume = "78",
    pages = "123009",
    year = "2008"
}

@article{Gerasimov:2021chj,
    author = "Gerasimov, Roman and Bhoj, Praneet and Kislat, Fabian",
    title = "{New Constraints on Lorentz Invariance Violation from Combined Linear and Circular Optical Polarimetry of Extragalactic Sources}",
    eprint = "2104.00238",
    archivePrefix = "arXiv",
    primaryClass = "astro-ph.CO",
    doi = "10.3390/sym13050880",
    journal = "Symmetry",
    volume = "13",
    number = "5",
    pages = "880",
    year = "2021"
}

@article{Friedman:2020bxa,
    author = "Friedman, Andrew S. and Gerasimov, Roman and Leon, David and Stevens, Walker and Tytler, David and Keating, Brian G. and Kislat, Fabian",
    title = "{Improved constraints on anisotropic birefringent Lorentz invariance and $CPT$ violation from broadband optical polarimetry of high redshift galaxies}",
    eprint = "2003.00647",
    archivePrefix = "arXiv",
    primaryClass = "astro-ph.HE",
    doi = "10.1103/PhysRevD.102.043008",
    journal = "Phys. Rev. D",
    volume = "102",
    number = "4",
    pages = "043008",
    year = "2020"
}

@article{Lewis:2019xzd,
 author         = "Lewis, Antony",
 title          = "{GetDist: a Python package for analysing Monte Carlo
                   samples}",
 year           = "2019",
 eprint         = "1910.13970",
 archivePrefix  = "arXiv",
 primaryClass   = "astro-ph.IM",
 SLACcitation   = "%%CITATION = ARXIV:1910.13970;%%",
 url            = "https://getdist.readthedocs.io"
}

@article{Harris:2020xlr,
    author = "Harris, Charles R. and others",
    title = "{Array programming with NumPy}",
    eprint = "2006.10256",
    archivePrefix = "arXiv",
    primaryClass = "cs.MS",
    doi = "10.1038/s41586-020-2649-2",
    journal = "Nature",
    volume = "585",
    number = "7825",
    pages = "357--362",
    year = "2020"
}

@article{Hunter:2007ouj,
    author = "Hunter, John D.",
    title = "{Matplotlib: A 2D Graphics Environment}",
    doi = "10.1109/MCSE.2007.55",
    journal = "Comput. Sci. Eng.",
    volume = "9",
    number = "3",
    pages = "90--95",
    year = "2007"
}

@article{Gubitosi:2009eu,
    author = "Gubitosi, Giulia and Pagano, Luca and Amelino-Camelia, Giovanni and Melchiorri, Alessandro and Cooray, Asantha",
    title = "{A Constraint on Planck-scale Modifications to Electrodynamics with CMB polarization data}",
    eprint = "0904.3201",
    archivePrefix = "arXiv",
    primaryClass = "astro-ph.CO",
    doi = "10.1088/1475-7516/2009/08/021",
    journal = "JCAP",
    volume = "08",
    pages = "021",
    year = "2009"
}

@article{Pagano:2009kj,
    author = "Pagano, Luca and de Bernardis, Paolo and De Troia, Grazia and Gubitosi, Giulia and Masi, Silvia and Melchiorri, Alessandro and Natoli, Paolo and Piacentini, Francesco and Polenta, Gianluca",
    title = "{CMB Polarization Systematics, Cosmological Birefringence and the Gravitational Waves Background}",
    eprint = "0905.1651",
    archivePrefix = "arXiv",
    primaryClass = "astro-ph.CO",
    doi = "10.1103/PhysRevD.80.043522",
    journal = "Phys. Rev. D",
    volume = "80",
    pages = "043522",
    year = "2009"
}

@article{Gubitosi:2011ue,
    author = "Gubitosi, Giulia and Migliaccio, Marina and Pagano, Luca and Amelino-Camelia, Giovanni and Melchiorri, Alessandro and Natoli, Paolo and Polenta, Gianluca",
    title = "{Using CMB data to constrain non-isotropic Planck-scale modifications to Electrodynamics}",
    eprint = "1106.6049",
    archivePrefix = "arXiv",
    primaryClass = "astro-ph.CO",
    doi = "10.1088/1475-7516/2011/11/003",
    journal = "JCAP",
    volume = "11",
    pages = "003",
    year = "2011"
}

@article{Gruppuso:2015xza,
    author = "Gruppuso, A. and Gerbino, M. and Natoli, P. and Pagano, L. and Mandolesi, N. and Melchiorri, A. and Molinari, D.",
    title = "{Constraints on cosmological birefringence from Planck and Bicep2/Keck data}",
    eprint = "1509.04157",
    archivePrefix = "arXiv",
    primaryClass = "astro-ph.CO",
    doi = "10.1088/1475-7516/2016/06/001",
    journal = "JCAP",
    volume = "06",
    pages = "001",
    year = "2016"
}

@article{Feng:2006dp,
    author = "Feng, Bo and Li, Mingzhe and Xia, Jun-Qing and Chen, Xuelei and Zhang, Xinmin",
    title = "{Searching for CPT Violation with Cosmic Microwave Background Data from WMAP and BOOMERANG}",
    eprint = "astro-ph/0601095",
    archivePrefix = "arXiv",
    reportNumber = "RESCEU-2-06",
    doi = "10.1103/PhysRevLett.96.221302",
    journal = "Phys. Rev. Lett.",
    volume = "96",
    pages = "221302",
    year = "2006"
}

@article{Gruppuso:2020kfy,
    author = "Gruppuso, Alessandro and Molinari, Diego and Natoli, Paolo and Pagano, Luca",
    title = "{Planck 2018 constraints on anisotropic birefringence and its cross-correlation with CMB anisotropy}",
    eprint = "2008.10334",
    archivePrefix = "arXiv",
    primaryClass = "astro-ph.CO",
    doi = "10.1088/1475-7516/2020/11/066",
    journal = "JCAP",
    volume = "11",
    pages = "066",
    year = "2020"
}

@article{Minami:2020odp,
    author = "Minami, Yuto and Komatsu, Eiichiro",
    title = "{New Extraction of the Cosmic Birefringence from the Planck 2018 Polarization Data}",
    eprint = "2011.11254",
    archivePrefix = "arXiv",
    primaryClass = "astro-ph.CO",
    doi = "10.1103/PhysRevLett.125.221301",
    journal = "Phys. Rev. Lett.",
    volume = "125",
    number = "22",
    pages = "221301",
    year = "2020"
}

@article{SPT:2020cxx,
    author = "Bianchini, F. and others",
    collaboration = "SPT",
    title = "{Searching for Anisotropic Cosmic Birefringence with Polarization Data from SPTpol}",
    eprint = "2006.08061",
    archivePrefix = "arXiv",
    primaryClass = "astro-ph.CO",
    reportNumber = "FERMILAB-PUB-20-264-AE",
    doi = "10.1103/PhysRevD.102.083504",
    journal = "Phys. Rev. D",
    volume = "102",
    number = "8",
    pages = "083504",
    year = "2020"
}

@article{Gluscevic:2012me,
    author = "Gluscevic, Vera and Hanson, Duncan and Kamionkowski, Marc and Hirata, Christopher M.",
    title = "{First CMB Constraints on Direction-Dependent Cosmological Birefringence from WMAP-7}",
    eprint = "1206.5546",
    archivePrefix = "arXiv",
    primaryClass = "astro-ph.CO",
    doi = "10.1103/PhysRevD.86.103529",
    journal = "Phys. Rev. D",
    volume = "86",
    pages = "103529",
    year = "2012"
}

@article{Contreras:2017sgi,
    author = "Contreras, Dagoberto and Boubel, Paula and Scott, Douglas",
    title = "{Constraints on direction-dependent cosmic birefringence from Planck polarization data}",
    eprint = "1705.06387",
    archivePrefix = "arXiv",
    primaryClass = "astro-ph.CO",
    doi = "10.1088/1475-7516/2017/12/046",
    journal = "JCAP",
    volume = "12",
    pages = "046",
    year = "2017"
}

@article{Bortolami:2022whx,
    author = "Bortolami, Marco and Billi, Matteo and Gruppuso, Alessandro and Natoli, Paolo and Pagano, Luca",
    title = "{Planck constraints on cross-correlations between anisotropic cosmic birefringence and CMB polarization}",
    eprint = "2206.01635",
    archivePrefix = "arXiv",
    primaryClass = "astro-ph.CO",
    doi = "10.1088/1475-7516/2022/09/075",
    journal = "JCAP",
    volume = "09",
    pages = "075",
    year = "2022"
}

@article{Lue:1998mq,
    author = "Lue, Arthur and Wang, Li-Min and Kamionkowski, Marc",
    title = "{Cosmological signature of new parity violating interactions}",
    eprint = "astro-ph/9812088",
    archivePrefix = "arXiv",
    reportNumber = "CU-TP-926, CAL-675",
    doi = "10.1103/PhysRevLett.83.1506",
    journal = "Phys. Rev. Lett.",
    volume = "83",
    pages = "1506--1509",
    year = "1999"
}

@book{fowles1989introduction,
  title={Introduction to Modern Optics},
  author={Fowles, G.R.},
  isbn={9780486659572},
  lccn={74023437},
  series={Dover Books on Physics Series},
  url={https://books.google.it/books?id=SL1n9TuJ5YMC},
  year={1989},
  publisher={Dover Publications}
}

@article{SimonsObservatory:2018koc,
    author = "Ade, Peter and others",
    collaboration = "Simons Observatory",
    title = "{The Simons Observatory: Science goals and forecasts}",
    eprint = "1808.07445",
    archivePrefix = "arXiv",
    primaryClass = "astro-ph.CO",
    doi = "10.1088/1475-7516/2019/02/056",
    journal = "JCAP",
    volume = "02",
    pages = "056",
    year = "2019"
}

@article{LiteBIRD:2022cnt,
    author = "Allys, E. and others",
    collaboration = "LiteBIRD",
    title = "{Probing Cosmic Inflation with the LiteBIRD Cosmic Microwave Background Polarization Survey}",
    eprint = "2202.02773",
    archivePrefix = "arXiv",
    primaryClass = "astro-ph.IM",
    month = "2",
    year = "2022"
}

@article{CMB-S4:2016ple,
    author = "Abazajian, Kevork N. and others",
    collaboration = "CMB-S4",
    title = "{CMB-S4 Science Book, First Edition}",
    eprint = "1610.02743",
    archivePrefix = "arXiv",
    primaryClass = "astro-ph.CO",
    reportNumber = "FERMILAB-FN-1024-A-AE",
    month = "10",
    year = "2016"
}

@article{ACT:2020frw,
    author = "Choi, Steve K. and others",
    collaboration = "ACT",
    title = "{The Atacama Cosmology Telescope: a measurement of the Cosmic Microwave Background power spectra at 98 and 150 GHz}",
    eprint = "2007.07289",
    archivePrefix = "arXiv",
    primaryClass = "astro-ph.CO",
    doi = "10.1088/1475-7516/2020/12/045",
    journal = "JCAP",
    volume = "12",
    pages = "045",
    year = "2020"
}

@article{Hu:2001bc,
    author = "Hu, Wayne and Dodelson, Scott",
    title = "{Cosmic Microwave Background Anisotropies}",
    eprint = "astro-ph/0110414",
    archivePrefix = "arXiv",
    reportNumber = "FERMILAB-PUB-01-328-A",
    doi = "10.1146/annurev.astro.40.060401.093926",
    journal = "Ann. Rev. Astron. Astrophys.",
    volume = "40",
    pages = "171--216",
    year = "2002"
}

@article{Montero-Camacho:2018vgs,
    author = "Montero-Camacho, Paulo and Hirata, Christopher M.",
    title = "{Exploring circular polarization in the CMB due to conventional sources of cosmic birefringence}",
    eprint = "1803.04505",
    archivePrefix = "arXiv",
    primaryClass = "astro-ph.CO",
    doi = "10.1088/1475-7516/2018/08/040",
    journal = "JCAP",
    volume = "08",
    pages = "040",
    year = "2018"
}

@article{Lemarchand:2018lfy,
    author = "Lemarchand, N. and Grain, J. and Hurier, G. and Lacasa, F. and Fert\'e, A.",
    title = "{Secondary CMB anisotropies from magnetized haloes - I. Power spectra of the Faraday rotation angle and conversion rate}",
    eprint = "1810.09221",
    archivePrefix = "arXiv",
    primaryClass = "astro-ph.CO",
    doi = "10.1051/0004-6361/201834485",
    journal = "Astron. Astrophys.",
    volume = "630",
    pages = "A149",
    year = "2019"
}

@article{De:2014qza,
    author = "De, Soma and Tashiro, Hiroyuki",
    title = "{Circular Polarization of the CMB: A probe of the First stars}",
    eprint = "1401.1371",
    archivePrefix = "arXiv",
    primaryClass = "astro-ph.CO",
    doi = "10.1103/PhysRevD.92.123506",
    journal = "Phys. Rev. D",
    volume = "92",
    number = "12",
    pages = "123506",
    year = "2015"
}

@article{Ejlli:2018ucq,
    author = "Ejlli, Damian",
    title = "{On the CMB circular polarization: I. The Cotton\textendash{}Mouton effect}",
    eprint = "1810.04947",
    archivePrefix = "arXiv",
    primaryClass = "astro-ph.CO",
    doi = "10.1140/epjc/s10052-019-6713-8",
    journal = "Eur. Phys. J. C",
    volume = "79",
    number = "3",
    pages = "231",
    year = "2019"
}

@article{Cooray:2002nm,
    author = "Cooray, Asantha and Melchiorri, Alessandro and Silk, Joseph",
    title = "{Is the cosmic microwave background circularly polarized?}",
    eprint = "astro-ph/0205214",
    archivePrefix = "arXiv",
    doi = "10.1016/S0370-2693(02)03291-4",
    journal = "Phys. Lett. B",
    volume = "554",
    pages = "1--6",
    year = "2003"
}

@article{Giovannini:2009ru,
    author = "Giovannini, Massimo",
    title = "{The V-mode polarization of the Cosmic Microwave Background}",
    eprint = "0909.3629",
    archivePrefix = "arXiv",
    primaryClass = "astro-ph.CO",
    reportNumber = "CERN-PH-TH-2009-171",
    doi = "10.1103/PhysRevD.80.123013",
    journal = "Phys. Rev. D",
    volume = "80",
    pages = "123013",
    year = "2009"
}

@article{Kosowsky:1996yc,
    author = "Kosowsky, Arthur and Loeb, Abraham",
    title = "{Faraday rotation of microwave background polarization by a primordial magnetic field}",
    eprint = "astro-ph/9601055",
    archivePrefix = "arXiv",
    reportNumber = "CFA-4273",
    doi = "10.1086/177751",
    journal = "Astrophys. J.",
    volume = "469",
    pages = "1--6",
    year = "1996"
}

@article{Alexander:2008fp,
    author = "Alexander, Stephon and Ochoa, Joseph and Kosowsky, Arthur",
    title = "{Generation of Circular Polarization of the Cosmic Microwave Background}",
    eprint = "0810.2355",
    archivePrefix = "arXiv",
    primaryClass = "astro-ph",
    doi = "10.1103/PhysRevD.79.063524",
    journal = "Phys. Rev. D",
    volume = "79",
    pages = "063524",
    year = "2009"
}

@article{Inomata:2018vbu,
    author = "Inomata, Keisuke and Kamionkowski, Marc",
    title = "{Circular polarization of the cosmic microwave background from vector and tensor perturbations}",
    eprint = "1811.04957",
    archivePrefix = "arXiv",
    primaryClass = "astro-ph.CO",
    reportNumber = "IPMU 18-0184",
    doi = "10.1103/PhysRevD.99.043501",
    journal = "Phys. Rev. D",
    volume = "99",
    number = "4",
    pages = "043501",
    year = "2019"
}

@article{Ejlli:2016avx,
    author = "Ejlli, Damian",
    title = "{Magneto-optic effects of the cosmic microwave background}",
    eprint = "1607.02094",
    archivePrefix = "arXiv",
    primaryClass = "astro-ph.CO",
    doi = "10.1016/j.nuclphysb.2018.08.003",
    journal = "Nucl. Phys. B",
    volume = "935",
    pages = "83--128",
    year = "2018"
}

@article{Bartolo:2019eac,
    author = "Bartolo, Nicola and Hoseinpour, Ahmad and Matarrese, Sabino and Orlando, Giorgio and Zarei, Moslem",
    title = "{CMB Circular and B-mode Polarization from New Interactions}",
    eprint = "1903.04578",
    archivePrefix = "arXiv",
    primaryClass = "hep-ph",
    doi = "10.1103/PhysRevD.100.043516",
    journal = "Phys. Rev. D",
    volume = "100",
    number = "4",
    pages = "043516",
    year = "2019"
}

@article{Ejlli:2017uli,
    author = "Ejlli, Damian",
    title = "{Millicharged fermion vacuum polarization in a cosmic magnetic field and generation of CMB elliptic polarization}",
    eprint = "1704.01894",
    archivePrefix = "arXiv",
    primaryClass = "hep-ph",
    doi = "10.1103/PhysRevD.96.023540",
    journal = "Phys. Rev. D",
    volume = "96",
    number = "2",
    pages = "023540",
    year = "2017"
}

@misc{SPT-Guidi,
   author = "Guidi, Federica",
    collaboration = "SPT",
    title = {Cosmology from SPT-3G},
    note = "\url{https://agenda.infn.it/event/28896/contributions/164821/attachments/91724/124648/11\%20-\%20Guidi\_CMBFerrara.pdf}",
    howpublished = "Conference: From Planck to the future of CMB, Ferrara, Italy ",
    month = "23-27 May",
    year = "2022"
}

@Misc{BK-Karkare,
    author = "Karkare,  Kirit S.",
    collaboration = "BICEP/KECK",
    title = {BICEP/Keck Array: B-Mode Polarization Results and Future Plans},
    note = "\url{https://agenda.infn.it/event/28896/contributions/164874/attachments/91596/124419/15\%20-\%20Karkare\%2020220526\_Karkare\_Ferrara\_aspresented.pdf}",
    howpublished = "Conference: From Planck to the future of CMB, Ferrara, Italy ",
    month = "23-27 May",
    year = "2022"
}

@Misc{ACT-Duivenvoorden,
    author = "Duivenvoorden, Adriaan",
    collaboration = "ACT",
    title = {The Atacama Cosmology Telescope: Science and Analysis pipeline},
    url = {https://agenda.infn.it/event/28896/contributions/164814/attahments/91720/124644/06\%20-\%20Adriaan\%20Duivenvoorden\%202022052\_ferrrara.pdf},
    howpublished = "Conference: From Planck to the future of CMB, Ferrara, Italy ",
    month = "23-27 May",
    year = "2022"
}

@article{Kahniashvili:2006dt,
    author = "Kahniashvili, Tina and Gogoberidze, Grigol and Ratra, Bharat",
    title = "{Gamma Ray Burst Constraints on Ultraviolet Lorentz Invariance Violation}",
    eprint = "astro-ph/0607055",
    archivePrefix = "arXiv",
    doi = "10.1016/j.physletb.2006.10.041",
    journal = "Phys. Lett. B",
    volume = "643",
    pages = "81--85",
    year = "2006"
}

@article{Tizchang:2016vef,
    author = "Tizchang, S. and Batebi, S. and Haghighat, M. and Mohammadi, R.",
    title = "{Cosmic microwave background polarization in non-commutative space-time}",
    eprint = "1605.09045",
    archivePrefix = "arXiv",
    primaryClass = "hep-ph",
    doi = "10.1140/epjc/s10052-016-4312-5",
    journal = "Eur. Phys. J. C",
    volume = "76",
    number = "9",
    pages = "478",
    year = "2016"
}

@article{Bavarsad:2009hm,
    author = "Bavarsad, E. and Haghighat, M. and Rezaei, Z. and Mohammadi, R. and Motie, I. and Zarei, M.",
    title = "{Generation of circular polarization of the CMB}",
    eprint = "0912.2993",
    archivePrefix = "arXiv",
    primaryClass = "hep-th",
    doi = "10.1103/PhysRevD.81.084035",
    journal = "Phys. Rev. D",
    volume = "81",
    pages = "084035",
    year = "2010"
}
